\documentclass[journal]{IEEEtran}

\usepackage[ruled, vlined]{algorithm2e}
\newcounter{ctr}
\newcommand{\stp}{\addtocounter{ctr}{1}\arabic{ctr}.}
\usepackage{alltt}
\usepackage{amsmath,amssymb,amsfonts}
\usepackage{amsthm}
\theoremstyle{definition}
\newtheorem{assumption}{Assumption}
\newtheorem{S1_assumption}{$\mathbb{S}_{1}$-Assumption}
\newtheorem{S2_assumption}{$\mathbb{S}_{2}$-Assumption}
\newtheorem{definition}{Definition}
\newtheorem{lemma}{Lemma}
\newtheorem{proposition}{Proposition}
\newtheorem{remark}{Remark}
\newtheorem{theorem}{Theorem}
\newtheorem{corollary}{Corollary}
\newtheorem{problem}{Problem}
\usepackage{balance}
\usepackage{bm}
\usepackage{breakurl}
\usepackage{caption}
\usepackage{cite}
\usepackage{color}
\usepackage{colortbl}
\usepackage{csquotes}
\usepackage{enumerate}
\usepackage{enumitem}
\setlist[description]{leftmargin=\parindent,labelindent=\parindent}
\usepackage{graphicx}
\usepackage{epstopdf} 
\usepackage{flushend}
\usepackage[hidelinks]{hyperref}
\usepackage[latin1]{inputenc}
\usepackage{makecell}
\usepackage{mathtools}
\usepackage{multirow}
\usepackage{pbox}
\usepackage{picinpar}
\usepackage{pifont}
\usepackage{siunitx}
\usepackage{soul}
\usepackage{subcaption}
\usepackage{textcomp}
\usepackage{txfonts}
\usepackage{url}

\begin{document}

\title{Emergency-Brake Simplex: Toward A Verifiably Safe Control-CPS Architecture for Abrupt Runtime Reachability Constraint Changes}
\author{Henghua Shen, \IEEEmembership{Member, IEEE}, Qixin Wang, \IEEEmembership{Member, IEEE}
\thanks{The first version of the paper was submitted on Jan 3, 2025. 
The research project related to this paper is supported in part by HK RGC T22-505/19-N (P0031331, RBCR, P0031259, RBCP),
PolyU 152002/18E (P0005550, Q67V), PolyU 152164/14E (P0004750, Q44B),  GRF 15207324 (P0051926, B-QCFM), 
G-PolyU503/16,    
by HKSAR Government and HKJCCT P0041424 (ZB5A), 
and by the HK PolyU fund P0042701 (CE09), 
P0046487 (CE0F), 
P0047916 (TACW), 
P0042699 (CE55), 
P0045578 (CE1C), 
P0043884 (CD6R), 
P0047965 (TAEB), 
P0047964 (TAEA), 
P0033695 (ZVRD),
P0013879 (BBWH), 
P0036469 (CDA8), 
P0043634 (1-TAB2),
P0043647 (1-TABF),
P0042721 (1-ZVG0),
LTG22-25/IICA/33 (TDG 2022-25), 
and TDG22-25/SMS-11.
}
\thanks{H. Shen was with the Dept. of Computing, The Hong Kong Polytechnic University, Hung Hom, HONG KONG SAR. He is now with Macau Millennium College (email: henghua.shen@dal.ca).}
\thanks{Q. Wang is with the Dept. of Computing, The Hong Kong Polytechnic University, Hung Hom, HONG KONG SAR (email: csqwang@polyu.edu.hk).}
} 



\maketitle


\begin{abstract}
When a system's constraints change abruptly, the system's reachability safety does no longer sustain. Thus, the system can reach a forbidden/dangerous value. 
Conventional remedy practically involves online controller redesign (OCR) to re-establish the reachability's compliance with the new constraints, which, however, is usually too slow. There is a need for an online strategy capable of managing runtime changes in reachability constraints. However, to the best of the authors' knowledge, this topic has not been addressed in the existing literature. 
In this paper, we propose a fast fault tolerance strategy to recover the system's reachability safety in runtime. Instead of redesigning the system's controller, we propose to change the system's reference state to modify the system's reachability to comply with the new constraints. 
We frame the reference state search as an optimization problem and employ the Karush-Kuhn-Tucker (KKT) method as well as the Interior Point Method (IPM) based Newton's method (as a fallback for the KKT method) for fast solution derivation.  The optimization also allows more future fault tolerance.
Numerical simulations demonstrate that our method outperforms the conventional OCR method in terms of computational efficiency and success rate. 
Specifically, the results show that the proposed method finds a solution $10^{2}$ (with the IPM-based Newton's method) $\sim 10^{4}$ (with the KKT method) times faster than the OCR method. Additionally, the improvement rate of the success rate of our method over the OCR method is $40.81\%$ without considering the deadline of run time. The success rate remains at $49.44\%$ for the proposed method, while it becomes $0\%$ for the OCR method when a deadline of $1.5 \; seconds$ is imposed.
\end{abstract}

%

%
{}
\definecolor{limegreen}{rgb}{0.2, 0.8, 0.2}
\definecolor{forestgreen}{rgb}{0.13, 0.55, 0.13}
\definecolor{greenhtml}{rgb}{0.0, 0.5, 0.0}

\section{Introduction}
\label{sect:Introduction}

Control \emph{Cyber-Physical Systems} (control-CPSs) are the inevitable results of the convergence of computing with control applications~\cite{sha2008}. A control-CPS consists of a physical subsystem (aka the ``\emph{plant}''), and a cyber subsystem. 

The \emph{plant's state} (aka ``\emph{plant state}'' or simply ``\emph{state}'') is typically represented as an $n$-dimensional vector, and the corresponding $n$-dimensional vector space is called the \emph{plant's state space} (or simply ``\emph{state space}''). 

The cyber subsystem can involve complicated software. Modern software can contain tens of thousands to over millions of lines of source code. It is well-known that software at this scale cannot be fully debugged~\cite{sommerville2015}. Yet many control-CPSs are safety critical, hence demand verifiable safety. This problem becomes even more significant with the rise of AI. Modern AI controller software may not only be buggy, but also unexplainable: hallucination may happen in unexpected circumstances.

To address this problem, the Simplex architecture is proposed~\cite{sha2001}. This architecture consists of two cyber subsystems. The first is a modern cyber subsystem (e.g. AI controller software), which is too complicated to be fully debugged/explained. The other is a conventional cyber subsystem, with simple linear controller and well-defined \emph{Lyapunov stability region}~\cite{brogan1991} in the plant's state space. 

During runtime, the modern cyber subsystem runs in the front, connecting the sensing input with the actuating output. The conventional cyber subsystem runs in the background, monitoring the plant state in real-time. Whenever the plant state reaches the border of the Lyapunov stability region, the conventional cyber subsystem immediately takes over the modern cyber subsystem, and steers the plant state back to the inner part of the Lyapunov stability region. The conventional cyber subsystem only returns the control back to the modern cyber subsystem when the plant state is sufficiently inside the Lyapunov stability region. 

In this way, the reachable plant state is guaranteed to be within the Lyapunov stability region of the conventional cyber subsystem. As long as this Lyapunov stability region never overlaps with unsafe states (collectively referred to as the ``\emph{forbidden region}'') in the plant's state space, the holistic control-CPS is verifiably safe, even if the modern cyber subsystem's behavior is unpredictable (due to bugs/unexplainability).

The forbidden region in the plant's state space is defined by a set of constraints, aka the \emph{reachability constraints}. The conventional Simplex architecture assumes the reachability constraints are given at the design stage. However, in practice, reachability constraint(s) can change in runtime, reshaping the forbidden region to overlap with the Lyapunov stability region. If this happens, the control-CPS is no longer verifiably safe.

One remedy is to redesign online the linear  controller of the conventional cyber subsystem (referred to as the \emph{Online-Controller-Redesign} (OCR) method), using the same design-stage procedures. However, such procedures are usually slow, and hence cannot give a redesigned controller and its Lyapunov stability region in time. Therefore, we need a fast enough alternative to cope with the runtime reachability constraint changes.

We propose not to redesign the linear controller of the conventional cyber subsystem. Instead, based on the present plant state, we change the \emph{reference state} (i.e. the targeted plant state) of the controller. This will immediately resize/move the Lyapunov stability region in the state space, to avoid the changed reachability constraints.

Specifically, we make the following contributions.

\begin{enumerate}
    \item We formulated the problem of dealing with runtime reachability constraint change as an \emph{Online Reference State Optimization Problem} (ORSOP). 
    \item We derived conditions under which the ORSOP has analytical solutions. 
    %
    \item When the analytical solution conditions do not sustain, we propose an \emph{Interior Point Method} (IPM) based numerical solution. 
    %
    \item We compare the performance of our ORSOP method with the OCR method under different computation time limits on our testbed. The ORSOP method can achieve a much higher success rate than the OCR method. Statistically, the ORSOP method can also achieve a much bigger safety margin than the OCR method.
\end{enumerate}

\section{Related Work}
\label{sect:RelatedWork}

The problem of preserving system safety in runtime has been studied in the fault-tolerant CPS literature. In what follows, we briefly review some closely relevant works and explain the differences.

\textbf{Model Checking}: Reachability has been a core concern in model checking that decides (during the design stage or runtime) if (starting from a given set of initial states) a forbidden region in the state space will be reached~\cite{xiang2020reachable}\cite{xiang2018reachability}\cite{li2014}. Thus, the focus of model checking is on finding proper approximations of a reachable set~\cite{kurzhanski2002reachability,tran2019safety,liu2023reachability,zhang2012safety}.
While this paper focuses on how to remedy the system in runtime, in case the runtime model checking alarms us that the forbidden region becomes reachable (due to runtime reachability constraint change). 

\textbf{Fallback Controller}: The Simplex architecture ~\cite{Simplex2001LuiSha,bak2014real,biondi2019safe} proposes to switch to a fallback high assurance controller in case of runtime (front end) controller failures. These works, however, do not cover runtime reachability constraint changes. In case of runtime reachability constraint changes, our paper's solution can complement the Simplex architecture by providing the needed high assurance controller, via simply changing the reference point.

\textbf{Plant Modification}: Another way to deal with runtime reachability constraint changes is to modify other parts of the system (typically, the plant) instead of the controller ~\cite{aseev2001optimal}\cite{MikhailGusev2015} (for example, discarding parts of the plant to change its physics). But this is not always feasible, and is not the focus of this paper.

\textbf{Path Re-Planning}: Some works on smart vehicles propose path re-planning in case of runtime reachability constraint changes~\cite{althoff2014online,schurmann2017ensuring,althoff2015online}. However, these works focus on simulating/analyzing one or a countable set of trajectories. While this paper focuses on the bound of all the possible trajectories.
In addition, the literature of~\cite{althoff2014online,schurmann2017ensuring,althoff2015online} assumes the plants are vehicles, while this paper assumes generic linear state-space models.

\section{Background}
\label{sect:Background}

\subsection{Control Theory}
\label{subsect:Control Theory}

In this paper, we focus on linear control systems, where the plant state at time $t$ is denoted as an $n$-dimensional vector\footnote{Unless otherwise specified, in this paper, a vector variable is denoted by a lower-case letter with an overhead arrow, while a scalar variable is denoted by a lower case letter without overhead arrow. A matrix variable is denoted by an upper-case letter.} $\vec{x}(t) =(x_1(t), x_2(t), \cdots, x_n(t))^{\sf T} \in \mathbb{R}^n$ (where $^{\sf T}$ means transpose). For simplicity, we also denote $\vec{x}(t)$ as $\vec{x}$, and denote the time derivative of $\vec{x}(t)$ as $\dot{\vec{x}}$.
%
%

Besides, the targeted plant state of the control, aka the \emph{reference state}, is denoted as $\vec{x}_{\sf o} \in \mathbb{R}^n$. We call the set of all feasible values for $\vec{x}_{\sf o}$ as the \emph{feasible region of the reference state}, denoted as $\mathcal{R}_{\sf o}$. In this paper, we assume the following.
\vspace{0.1in}
\hrule
\begin{assumption}
\label{assumption:ClosedFeasibleRegionOfTheReferenceState}
$\mathcal{R}_{\sf o}$ is \emph{closed}, and is defined by a set of linear constraints, aka \emph{reference state constraints}, denoted by 
\begin{eqnarray}
    \label{eqn:ReferenceStateConstraints}
    g_j(\vec{x}_{\sf o}) \stackrel{{\sf def}}{=} \vec{\omega}_j \cdot \vec{x}_{\sf o} + b_j \leqslant 0, \;\; j = 1, 2, \ldots, r. 
\end{eqnarray}
\end{assumption}
\hrule
\vspace{0.1in}
\hrule
\begin{assumption}
\label{assumption:ConstantReferenceState}
    Unless otherwise denoted (specifically,  when switching the reference state), we assume $\vec{x}_{\sf o}$ is constant. 
\end{assumption}
\hrule
\vspace{0.1in}

With the above notations, the dynamics of a \emph{linear time-invariant control system} (simplified as ``\emph{linear control system}'' in the following) is described by
\begin{eqnarray}
    \begin{cases}
        \dot{\vec{x}} = A(\vec{x} - \vec{x}_{\sf o})+B\vec{u}, \\
        \vec{u} = -K(\vec{x} - \vec{x}_{\sf o}), 
    \end{cases} 
    \label{eqn:LinearSystemModel}
\end{eqnarray}
where $A \in \mathbb{R}^{n \times n}$ and $B \in \mathbb{R}^{n \times m}$ are constant matrices; $\vec{u} \in \mathbb{R}^m$ is the control signal outputted by the \emph{linear controller} $\vec{u}=-K(\vec{x} - \vec{x}_{\sf o})$; and $K\in \mathbb{R}^{m\times n}$ is the constant \emph{controller matrix}. 

\vspace{0.1in}
\hrule
\begin{definition}
    \label{def:GAS}
    The linear control system \eqref{eqn:LinearSystemModel} is \emph{Globally Asymptopitcally Stable} (GAS) iff starting from any $\vec{x}(t_0) \in \mathbb{R}^n$ (where $t_0$ is the initial time instance), the trajectory of $\vec{x}(t) \rightarrow \vec{x}_{\sf o}$ as $t \rightarrow +\infty$.
\end{definition}
\hrule
\vspace{0.1in}

We have the following well-known lemma~\cite{brogan1991}\cite{brogan1985modern}.

\vspace{0.1in}
\hrule
\begin{lemma}
    \label{lemma:GAS}
    Given the linear control system \eqref{eqn:LinearSystemModel} (where $\vec{x}_{\sf o}$ is a given constant). Suppose the following condition {\bf C1} sustains. 
    \begin{description}
        \item[({\bf C1}):] There exist constant symmetric positive definite matrices $P \in \mathbb{R}^{n \times n}$ and $Q \in \mathbb{R}^{n \times n}$, which solve the \emph{Lyapunov equation} 
        \begin{eqnarray}
        A_{\sf cl}^{\sf T}P + PA_{\sf cl} = -Q, \label{eqn:LyapunovEquation}
        \end{eqnarray}
        where $A_{\sf cl} \stackrel{{\sf def}}{=} (A-BK) \in \mathbb{R}^{n\times n}$.
    \end{description}
    Then we have the following.
    \begin{enumerate}
        \item Denote \emph{Lyapunov function} 
            \begin{eqnarray}
                V_{\vec{x}_{\sf o}, P}(\vec{x}) \stackrel{{\sf def}}{=} (\vec{x} - \vec{x}_{\sf o})^{\sf T}P(\vec{x} - \vec{x}_{\sf o}), \label{eqn:LyapunovFunction}
            \end{eqnarray}
            we have $\forall \vec{x} \in \mathbb{R}^n$, $V_{\vec{x}_{\sf o}, P}(\vec{x}) \geqslant 0$; and $V_{\vec{x}_{\sf o}, P}(\vec{x})=0$ iff $\vec{x}=\vec{x}_{\sf o}$.

        \item The linear control system \eqref{eqn:LinearSystemModel} is GAS.
        \item $\forall \vec{x} \in \mathbb{R}^n$, $\dot{V}_{\vec{x}_{\sf o}, P}(\vec{x}) \leqslant 0$; and $\dot{V}_{\vec{x}_{\sf o}, P} (\vec{x}) = 0$ iff $\vec{x} = \vec{x}_{\sf o}$. 
    \end{enumerate}
\end{lemma}
\hrule
\vspace{0.1in}


If condition {\bf C1} in Lemma~\ref{lemma:GAS} sustains, given the initial plant state of $\vec{x}(t_0)$, then Lemma~\ref{lemma:GAS} basically says that the future trajectory of
$\vec{x}(t)$ ($t \geqslant t_0$), denoted as $\{\vec{x}(t)\}_{t \geqslant t_{0}}$, is confined by the hyper ellipsoid, 
%
aka \emph{Lyapunov ellipsoid}, of 
\begin{eqnarray}
    \label{eqn:LyapunovEllipsoidFunction}
    E\left(\vec{x}(t_0), \vec{x}_{\sf o}, P\right) \stackrel{{\sf def}}{=} \left\{ \vec{\xi} \, \big| \, \vec{\xi} \in \mathbb{R}^n \mbox{ and }  V_{\vec{x}_{\sf o}, P}(\vec{\xi}) \leqslant V_{\vec{x}_{\sf o}, P}(\vec{x}(t_0)) \right\},  
\end{eqnarray}
where intuitively, $\vec{x}_{{\sf o}}$ decides the center of the hyper ellipsoid, $P$ decides the shape and orientation of the hyper ellipsoid, and $\vec{x}(t_0)$, as a point on the surface,  decides (together with $\vec{x}_{{\sf o}} $ and $P$) the size of the hyper ellipsoid.
\noindent The Lyapunov ellipsoid $E\left(\vec{x}(t_0), \vec{x}_{\sf o}, P\right)$ bounds the \emph{reachable region} of the plant state $\vec{x}$ of the linear control system \eqref{eqn:LinearSystemModel}, given the initial plant state $\vec{x}(t_0)$. In this sense, the Lyapunov ellipsoid is a so-called \emph{Lyapunov stability region}~\cite{brogan1991}\cite{brogan1985modern}. \emph{In the following, unless otherwise denoted, we use the term ``Lyapunov ellipsoid'' and ``Lyapunov stability region'' interchangeably}.  

Meanwhile, a linear control system \eqref{eqn:LinearSystemModel} often has to guarantee the \emph{reachability safety}. Specifically, the plant state $\vec{x}$ can never enter a set of \emph{forbidden region(s)}, collectively denoted as $\mathcal{F} \subseteq \mathbb{R}^n$. Usually, $\mathcal{F}$ is determined by safety concerns and plant's physical constraints. Mathematically, these constraints are specified by a set of linear/non-linear inequalities, collectively called the ``\emph{reachability constraints}.'' 
For narrative simplicity, we call $\bar{\mathcal{F}} \stackrel{{\sf def}}{=} \mathbb{R}^n - \mathcal{F}$ the \emph{operational region(s)}, and the corresponding linear/non-linear inequalities that define $\bar{\mathcal{F}}$ the ``\emph{operational constraints}.'' 
In this paper, we focus on the cases where all operational constraints are linear, and $\bar{\mathcal{F}}$ is compact (i.e. closed and bounded) and convex (see Assumption~\ref{assumption:LinearConstraints}). Meanwhile, as $\bar{\mathcal{F}}$ and $\mathcal{F}$ imply each other, operational constraints and reachability constraints also imply each other. For narrative simplicity, in the following, we may either use ``operational constraints'' or ``reachability constraints'' depending on the context.

\begin{figure}[h]
    \centering  
    \includegraphics[scale=0.33, trim = 3.5cm 1.5cm 6cm 0.5cm, clip]{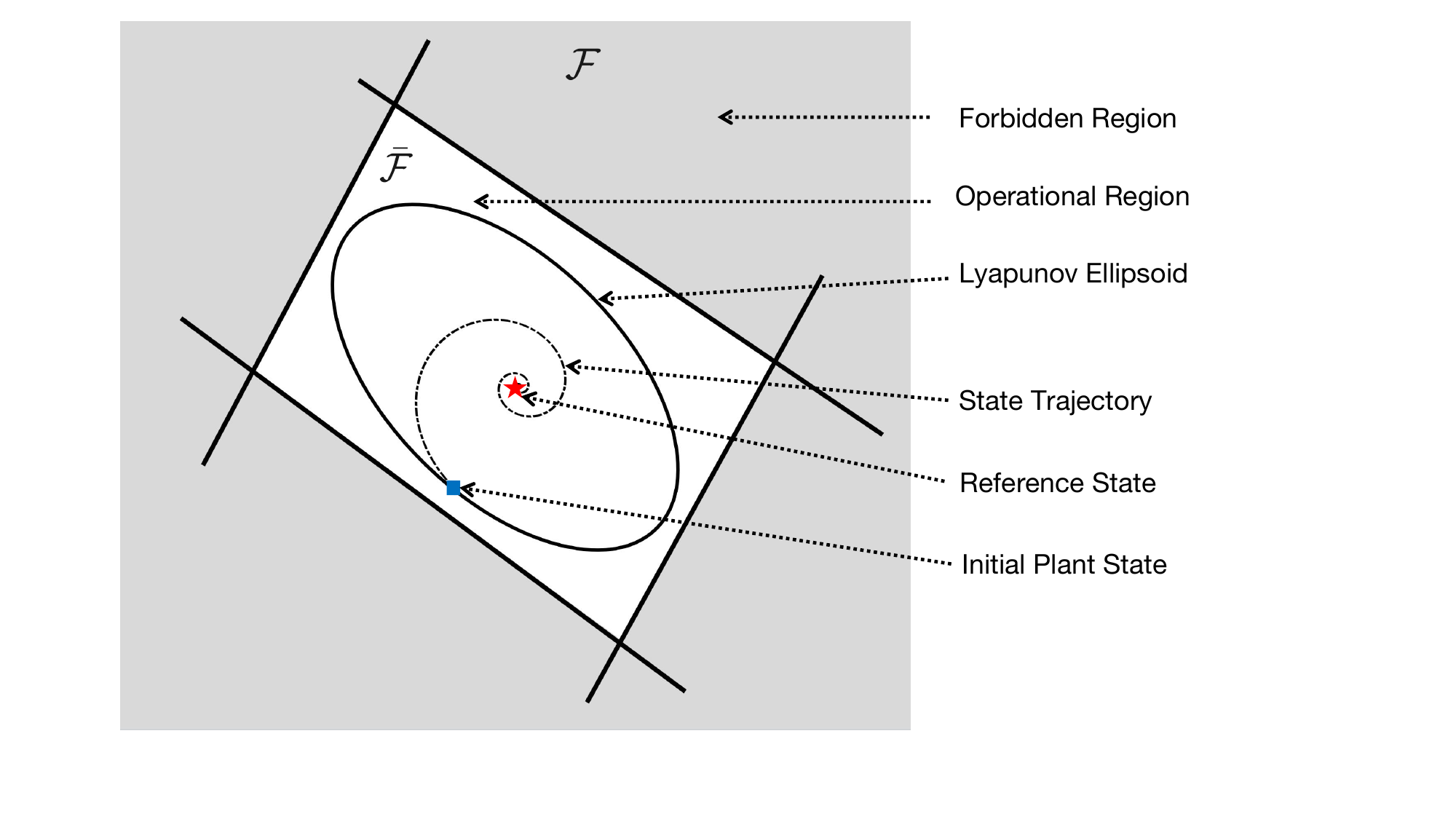}  
    \caption{Illustration of Lyapunov ellipsoid, forbidden region, operational region, initial state, state trajectory, and reference state.}  
    \label{fig:IllustrationOfRegionsAndBounds}   
\end{figure}

Fig.~\ref{fig:IllustrationOfRegionsAndBounds} illustrates the concepts of Lyapunov ellipsoid, forbidden region, operational region, initial plant state, state trajectory, and reference state.


\subsection{KKT Method}
\label{subsect:KktOptimization}

In convex optimization, the KKT conditions~\cite{kuhn2013nonlinear} are a set of necessary conditions for the optimal solution(s), which is described as follows:

\vspace{0.1in}
\hrule
\begin{lemma} 
    \label{lemma:KKT}
    Given a convex optimization problem of the form:
    \begin{eqnarray}
        && {\sf min}_{\vec{x}} \;\; f(\vec{x}),  \label{eqn:ObjectiveProblemKKTMethodology} \\
        \mbox{s.t. } && f_i(\vec{x})\leqslant 0, \; i=1, 2, \ldots, h, \label{eqn:ProblemKKTMethodologyConstraints} \\
        && \vec{x} \in \mathbb{R}^n. \label{eqn:ProblemKKTMethodologyDomain}
    \end{eqnarray}
    Assume that $f({\vec{x}})$ and $f_i(\vec{x})$ ($i = 1$, $2$, $\ldots$, $h$) are convex and differentiable. Then the corresponding \emph{Lagrangian function} is defined as
    \begin{eqnarray}
        L(\vec{x}, \vec{\mu}) \stackrel{{\sf def}}{=} f(\vec{x}) + \sum_{i=1}^{h} \mu_i f_i(\vec{x}), \label{eqn:L_x}
    \end{eqnarray}
    \noindent where $\vec{\mu} \stackrel{{\sf def}}{=} (\mu_1$, $\mu_2$, $\ldots$, $\mu_h)^{{\sf T}} \in \mathbb{R}^h$ is the so-called \emph{Lagrange multiplier} vector. 
    Denote the optimal solution to \eqref{eqn:ObjectiveProblemKKTMethodology} as $\vec{x}^*$. If $\vec{x}^*$ exists, then there exists $\vec{\mu}^*=(\mu^*_1$, $\mu^*_2$, $\dots$, $\mu^*_h)^{{\sf T}} \in \mathbb{R}^h$ such that the following conditions (aka \emph{KKT conditions}) sustain:
    \begin{enumerate}
        \item Stationarity: $\frac{\partial L(\vec{x}^*, \vec{\mu}^*)}{\partial \vec{x}} = \mathbf{0}$, i.e.  $\frac{\partial f(\vec{x}^*)}{\partial \vec{x}} + \sum_{i=1}^{h} \mu^*_i \frac{\partial f_i(\vec{x}^*)}{\partial \vec{x}} = \mathbf{0}$;
        \item Primal Feasibility: $\vec{x}^* \in \mathbb{R}^n$, and $f_i(\vec{x}^*) \leqslant 0$ ($i = 1$, $2$, $\ldots$, $h$);
        \item Dual Feasibility: $\mu^*_i \geqslant 0$ ($i = 1$, $2$, $\ldots$, $h$);
        \item Complementary Slackness: $\mu^*_i f_i(\vec{x}^*)=0$, 
        ($i = 1$, $2$, $\ldots$, $h$).
    \end{enumerate}
\end{lemma}
\hrule
\vspace{0.1in}

Lemma~\ref{lemma:KKT} establishes a set of necessary conditions (aka KKT conditions) for any optimal solution $\vec{x}^*$ to \eqref{eqn:ObjectiveProblemKKTMethodology}. Often we can analytically derive the set of all solutions $\mathcal{S}$ that meet these necessary conditions. Any optimal solution $\vec{x}^*$ to \eqref{eqn:ObjectiveProblemKKTMethodology} should then belong to $\mathcal{S}$. In case $\mathcal{S}$ is enumerable, then by checking $\mathcal{S}$'s elements individually, we can find $\vec{x}^*$.

\subsection{Newton's Method}
\label{subsect:NewtonsMethod}

The KKT method mentioned in Section~\ref{subsect:KktOptimization} to find $\vec{x}^*$ is analytical. However, this analytical method is not guaranteed to work in all situations, especially when the constraint \eqref{eqn:ProblemKKTMethodologyConstraints} is highly nonlinear. Alternatively, we can try the numerical \emph{unconstrained Newton's method} (simplified as the ``\emph{Newton's method}'' in the following), which iteratively searches for a solution for a given unconstrained optimization problem: 
\begin{eqnarray}
{\sf min}_{\vec{x}} \;\; F(\vec{x}),  \mbox{ where } \vec{x} \in \mathbb{R}^n. \label{eqn:UnconstrainedOptimizationProblem} 
\end{eqnarray}

The iteration formula is
\begin{eqnarray}
    \label{eqn:NewtonIterationFunction}
    \vec{x}^{(\imath+1)} = \vec{x}^{(\imath)}-\eta^{(\imath)}{[\nabla^2 F(\vec{x}^{(\imath)})]}^{-1} {\nabla F(\vec{x}^{(\imath)})}, 
\end{eqnarray}
where $\imath$ indexes the iteration; $\nabla F(\vec{x})$ is the gradient of $F(\vec{x})$; and $\nabla^2 F(\vec{x})$ is the Hessian matrix of $F(\vec{x})$. The step size at $\imath$th iteration is denoted by $\eta^{(\imath)} >0$, which can be fixed or adaptive~\cite{polyak2020new}. The iteration of  \eqref{eqn:NewtonIterationFunction} repeats until one of the following ending conditions sustains:
\begin{description}
\item[({\bf E1}):] The error $\|\vec{x}^{(\imath+1)} - \vec{x}^{(\imath)}\|_2$ (where $\|\cdot\|_2$ is the Euclidean norm) converges within a predefined small enough bound $\varepsilon > 0$, and $|F(\vec{x}^{(\imath+1)})| < +\infty$.
\item[({\bf E2}):] A maximum iteration count $n_{\sf max}$ is hit.
\end{description}

In the case of {\bf E1}, we claim the solution to the  optimization problem \eqref{eqn:UnconstrainedOptimizationProblem} is found: $\vec{x}^*=\vec{x}^{(\imath+1)}$. Otherwise, we claim ``failure.''

To convert the constrained optimization problem \eqref{eqn:ObjectiveProblemKKTMethodology}\eqref{eqn:ProblemKKTMethodologyConstraints}\eqref{eqn:ProblemKKTMethodologyDomain} to an unconstrained optimization problem of form \eqref{eqn:UnconstrainedOptimizationProblem}, the ``\emph{Barrier Method},'' aka ``\emph{Interior-Point Method} (IPM),'' is commonly used~\cite{wright2001convergence}. 

IPM needs an indicator function  
%
\begin{eqnarray}
    \label{eqn:IndicatorFunction}
    \mathbb{I}(\xi) \stackrel{{\sf def}}{=} \begin{cases}
        \; 0, \;\; \mbox{ if } \;\; \xi \leqslant 0; \\
        \; +\infty, \;\; \mbox{if } \;\; \xi > 0.

    \end{cases}
\end{eqnarray}
However, the above $\mathbb{I}(\xi)$ is not differentiable, hence is inconvenient to use.  A popular solution is to use the natural logarithm function ${\sf ln}(\cdot)$ to approximate the indicator function as follows:
\begin{eqnarray}
    \label{eqn:IndicationFunctionApproximate_Log}
    \mathbb{I}(\xi) \approx - \frac{1}{\lambda} {\sf ln} (-\xi),  
\end{eqnarray}
where $\lambda>0$ is a large enough number (e.g., $\lambda=10^6$~\cite{ghojogh2021kkt}) and larger $\lambda$ allows for a more accurate approximation~\cite[pp.563]{boyd2004convex}.
Then, the constrained optimization problem \eqref{eqn:ObjectiveProblemKKTMethodology}\eqref{eqn:ProblemKKTMethodologyConstraints}\eqref{eqn:ProblemKKTMethodologyDomain} is converted to the following unconstrained form: 
\begin{eqnarray}
    \label{eqn:UnconstrainedObjectiveFunction4NewtonMethod}
     {\sf min}_{\vec{x}} \left(F(\vec{x}) \stackrel{{\sf def}}{=} f(\vec{x})- \frac{1}{\lambda} \sum _{i=1}^h {\sf ln} (-f_i(\vec{x}))\right), \mbox{ where } \vec{x} \in \mathbb{R}^{n}, 
\end{eqnarray}
\noindent which can be solved using the unconstrained Newton's method described by \eqref{eqn:NewtonIterationFunction}.

Note there is still an implementation issue to take care of. ${\sf ln}(-\xi)$ is undefined when $\xi \geqslant 0$. Correspondingly, \eqref{eqn:IndicationFunctionApproximate_Log} is undefined when $\xi \geqslant 0$, and $F(\vec{x})$ of \eqref{eqn:UnconstrainedObjectiveFunction4NewtonMethod} is undefined when $f_i(\vec{x}) \geqslant 0$ ($i \in \{1, \dots, h\}$). In practice, in each iteration step $\imath \in \mathbb{N}$, we need to check this. Specifically, if $\exists i \in \{1, \ldots, h\}$, s.t. $f_i(\vec{x}^{(\imath)}) \geqslant 0$, we will stop the iteration and claim the failure of the IPM-based Newton's method. In other words, $\forall \imath \in \mathbb{N}$, we need to assert 
\begin{eqnarray}
    \forall i \in \{1, \ldots, h\}, \quad f_i(\vec{x}^{(\imath)}) < 0; \label{eqn:IPMPerStepAssert}
\end{eqnarray}
\noindent otherwise, we need to stop the iteration and claim the failure of the IPM-based Newton's method. $\hfill (*)$

\section{Problem Formulation}
\label{sect:ProblemFormulation}

The Simplex architecture~\cite{sha2001} assumes the conventional cyber subsystem to be a linear control system of \eqref{eqn:LinearSystemModel}.

Given \eqref{eqn:LinearSystemModel} and the forbidden region $\mathcal{F}$ (defined by a set of reachability constraints), where $A$ and $B$ are known, there are mature routines (e.g. the seminal LMI method~\cite{boyd1994linear}) to numerically find $K$, $P$, and $Q$, such that ({\bf C1}) of Lemma~\ref{lemma:GAS} sustains, which also results in a Lyapunov ellipsoid $\mathcal{E} = E(\vec{x}(t_0), \vec{x}_{\sf o}, P)$ (see \eqref{eqn:LyapunovEllipsoidFunction}), such that $\mathcal{E} \cap \mathcal{F} = \varnothing$. As the trajectory of the plant state $\{\vec{x}(t)\}_{t \geqslant t_{0}}$ is confined by the Lyapunov ellipsoid $\mathcal{E}$ (i.e. $\{\vec{x}(t)\}_{t \geqslant t_{0}} \subseteq \mathcal{E}$), so we have $\{\vec{x}(t)\}_{t \geqslant t_{0}} \cap \mathcal{F} = \varnothing$. That is, the linear control system \eqref{eqn:LinearSystemModel} guarantees the  reachability safety.

\begin{figure}[]
    \centering  
    \includegraphics[scale=0.3, trim = 3.5cm 0.1cm 0.2cm 0.2cm, clip]{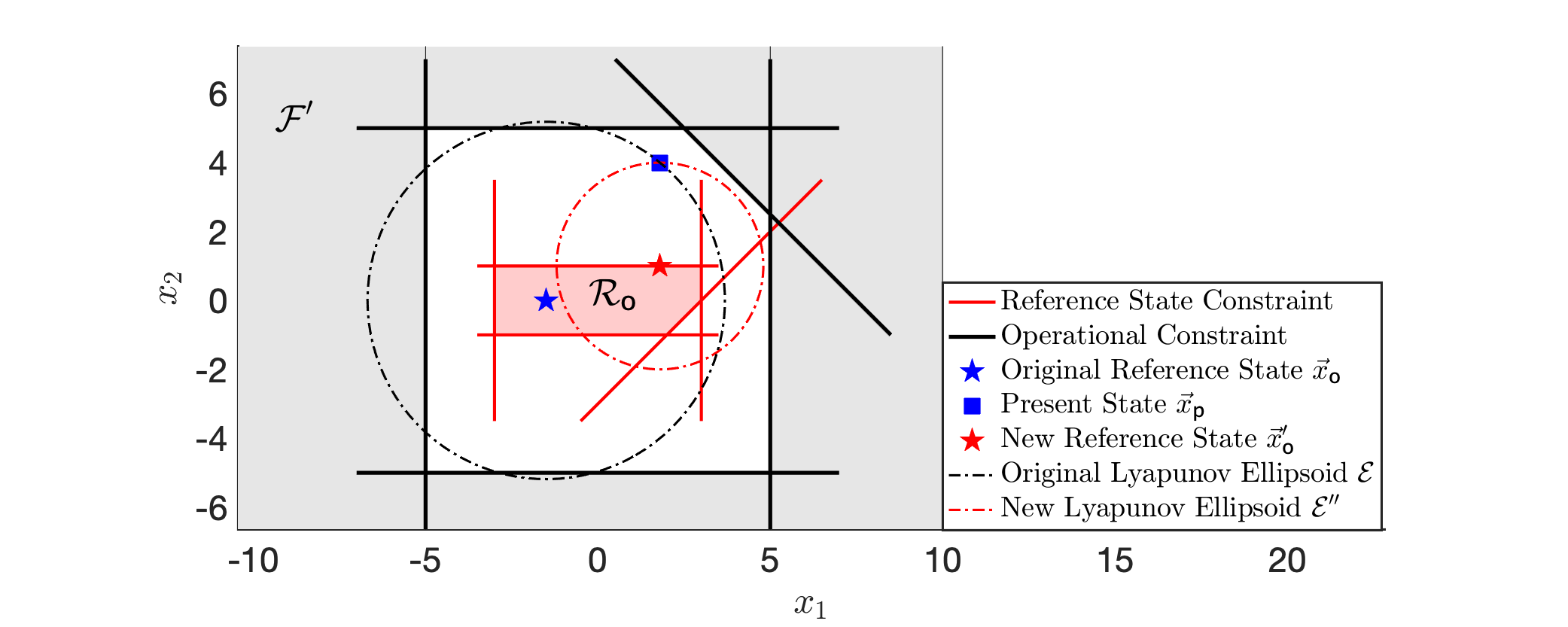}  
    \caption{Illustration of an original Lyapunov ellipsoid $\mathcal{E}$ (delineated by the black dash-dot line) violating the new reachability constraints (delineated by the black solid lines) in 2D space. We intend to find a new reference state $\vec{x}'_{\sf o}$ (marked by the red star) in the feasible region of the reference state $\mathcal{R}_{\sf o}$ (the red area delineated by the red solid lines), so that the new Lyapunov ellipsoid $\mathcal{E}''$ (delineated by the red dash-dot line) does not overlap with the new forbidden region $\mathcal{F}'$ (the gray area delineated by the black solid lines).} 
    \label{fig:IllustrationOfFindingNewLyapunovEllipse}   
\end{figure}

However, the above assumes the forbidden region $\mathcal{F}$ never changes. As illustrated in Fig.~\ref{fig:IllustrationOfFindingNewLyapunovEllipse} (in 2D space as an example), if $\mathcal{F}$ changes to $\mathcal{F}'$ at time instance $t_1$ ($t_1 > t_0$), then $\mathcal{E}$ (delineated by the black dash-dot line) may overlap with $\mathcal{F}'$, i.e. $\mathcal{E} \cap \mathcal{F}' \neq \varnothing$, breaking the guarantee of reachability safety.

As described in Section~\ref{sect:Introduction}, the conventional remedy is to carry out the \emph{Online-Controller-Redesign} (OCR), i.e. to redesign the controller online, to derive the new $K'$, $P'$, $Q'$, and $\mathcal{E}'=E(\vec{x}(t_1), \vec{x}_{\sf o}, P')$, so that ({\bf C1}) of Lemma~\ref{lemma:GAS} sustains, and $\mathcal{E}' \cap \mathcal{F}' = \varnothing$.

However, often the reachability safety guarantee needs to be recovered in real-time. OCR incurs controller redesign, which costs too much time. To meet the real-time demand, \emph{we propose only to find a new reference state $\vec{x}'_{\sf o}$} (see the red star in Fig.~\ref{fig:IllustrationOfFindingNewLyapunovEllipse}), while keep all other parts of the \emph{original linear control system} \eqref{eqn:LinearSystemModel} (particularly, the original controller matrix $K$) unchanged. 

\vspace{0.1in} 

That is, the \emph{new linear control system} becomes
\begin{eqnarray}
\begin{cases}
\dot{\vec{x}} = A(\vec{x} - \vec{x}'_{\sf o})+B\vec{u}, \\
\vec{u} = -K(\vec{x} - \vec{x}'_{\sf o}), 
\end{cases} 
\label{eqn:NewLinearSystemModel}
\end{eqnarray}

We demand $\vec{x}'_{\sf o}$ to satisfy the following requirements. 
%
%
\begin{description}
    \item[({\bf R1}):] (Obligatory) Confine the new linear control system \eqref{eqn:NewLinearSystemModel}'s future trajectory of $\vec{x}(t)$ ($t \geqslant t_{1}$), denoted as $\{\vec{x}(t)\}_{t \geqslant t_{1}}$, within a new Lyapunov ellipsoid of the following form
    \begin{eqnarray}
    \label{eqn:NewLyapunovEllipsoid}
    && \mathcal{E}'' = E(\vec{x}(t_1), \vec{x}'_{\sf o}, P) \nonumber \\
    &=& \left\{ \vec{\xi} \, \big| V_{\vec{x}'_{\sf o}, P} (\vec{\xi}) \leqslant V_{\vec{x}'_{\sf o}, P}(\vec{x}(t_1)),\;\vec{\xi} \in \mathbb{R}^n \right\}, 
    \end{eqnarray}
    \noindent where (in compliance with the definition by  \eqref{eqn:LyapunovFunction}) 
    \begin{eqnarray}
    \label{eqn:NewLyapunovPotentialEnergy}
    V_{\vec{x}'_{\sf o}, P}(\vec{\xi}) = (\vec{\xi} - \vec{x}'_{\sf o})^{{\sf T}} P (\vec{\xi} - \vec{x}'_{\sf o}), 
    \end{eqnarray}
    \noindent and $\mathcal{E}'' \cap \mathcal{F}' = \varnothing$. For example, in Fig.~\ref{fig:IllustrationOfFindingNewLyapunovEllipse}, $\mathcal{E}''$ (the shape delineated by the red dash-dot line) should not overlap with the new forbidden region $\mathcal{F}'$ (the gray area delineated by the black solid lines). Note $\{\vec{x}(t)\}_{t \geqslant t_{1}} \subseteq \mathcal{E}''$ (i.e. confinement of $\{\vec{x}(t)\}_{t \geqslant t_{1}}$ in $\mathcal{E}''$), hence $\mathcal{E}'' \cap \mathcal{F}' = \varnothing$ implies $\{\vec{x}(t)\}_{t \geqslant t_{1}} \cap \mathcal{F}' = \varnothing$, i.e. the new linear control system's reachabiilty safety is guaranteed.
    \item[({\bf R2}):] (Obligatory) Confine $\vec{x}'_{\sf o}$ within the feasible region of the reference state (see  \eqref{eqn:ReferenceStateConstraints}), i.e. $\vec{x}'_{\sf o} \in \mathcal{R}_{\sf o}$. For example, in Fig.~\ref{fig:IllustrationOfFindingNewLyapunovEllipse}, the new reference state $\vec{x}'_{\sf o}$ (marked by the red star) should reside in the feasible region of the reference state $\mathcal{R}_{\sf o}$ (the red area delineated by the red solid lines).
    \item[({\bf R3}):] (Optional and Heuristic) Minimize the volume of $\mathcal{E}''$.
\end{description}
\vspace{0.1in}

Requirement {\bf R1} and {\bf R2} are obligatory. As long as they are satisfied, the plant state's reachability safety under the new reachability constraints is  guaranteed. 
Requirement {\bf R3} is optional and heuristic: minimizing the volume of $\mathcal{E}''$ makes $\mathcal{E}''$ more tolerant to further changes of the reachability constraints. 

To find $\vec{x}'_{\sf o}$ meeting the above requirements, let us clarify some more assumptions.

First, in this paper, we focus on linear operational constraints, which define compact and convex operational regions. Formally, we have 
\vspace{0.1in}
\hrule
\begin{assumption} 
    \label{assumption:LinearConstraints}
    The new operational region $\bar{\mathcal{F}}'$ is compact (i.e. closed and bounded) and convex, and is defined by a set of linear operational constraints:
    \begin{eqnarray}
        \label{eqn:NewReachabilityConstraints}
        \vec{v}_k \cdot \vec{x} + \beta_k \leqslant 0,\;\; k=1,2,\ldots, s. 
    \end{eqnarray}
\end{assumption}
\hrule
\vspace{0.1in}

Second, the present plant state $\vec{x}(t_1)$ must be in the operational region $\bar{\mathcal{F}}'$; for otherwise, there is no way to rescue the plant. Formally, we have
\vspace{0.1in}
\hrule
\begin{assumption} 
    \label{assumption:Rescueable}
    The present plant state $\vec{x}(t_1) \in \bar{\mathcal{F}}'$.
\end{assumption}
\hrule
\vspace{0.1in}

Third, for the time being, we further assume the following (note we will remove Assumption~\ref{assumption:PMatrixEqualEigenValue} in Section~\ref{subsect:RelaxationOfAssumption3}):
\vspace{0.1in}
\hrule
\begin{assumption} 
    \label{assumption:PMatrixEqualEigenValue}
    The original Lyapunov ellipsoid $\mathcal{E} = E(\vec{x}(t_0), \vec{x}_{\sf o}, P)$ of the original controller $\vec{u} = -K (\vec{x} - \vec{x}_{\sf o})$ has equal principal axes lengths (i.e. $\mathcal{E}$ is a hyper sphere). In other words, all the eigenvalues of $P$ have a same positive real value. 
\end{assumption}
\hrule
\vspace{0.1in}

Also, for narrative convenience, in the following, we denote the present plant state $\vec{x}(t_1)$ as $\vec{x}_{\sf p} =(x_{{\sf p}1},x_{{\sf p}2},\ldots,x_{{\sf p}n})^{\sf T} \in \mathbb{R}^n$. 

With the above assumptions and notations, the search for a new reference state $\vec{x}'_{\sf o}$ satisfying requirement {\bf R1} $\sim$ {\bf R3} can be formulated as Problem~\ref{Problem_I}, aka the \emph{Online Reference State Optimization Problem} (ORSOP). The reason why the ORSOP's solution satisfies requirement {\bf R1} $\sim$ {\bf R3} 
will be explained by Theorem~\ref{Theorem:ConfinementOfNewTrajectory} and Corollary~\ref{Corollary:ValidityOfOrsop}. 

\vspace{0.1in}
\hrule
\begin{problem}[ORSOP]
    \label{Problem_I}
    \begin{eqnarray}
        &&{\sf min}_{\vec{x}'_{\sf o}}    \; \;  \left( f(\vec{x}'_{\sf o}) \stackrel{{\sf def}}{=} \|\vec{x}'_{\sf o}-\vec{x}_{\sf p}\|_2^2 \right), \label{eqn:ObjectiveFunctionReachability_Problem1}\\
        \mbox{s.t.} \;\;\;\;\;  && g_j(\vec{x}'_{\sf o}) \stackrel{{\sf def}}{=} \vec{\omega}_j\cdot  \vec{x}'_{\sf o}+b_j \leqslant 0,\;\; j=1,2,\ldots, r; \qquad \quad \label{eqn:LinearConstraints_Problem1}\\
        &&  q_k(\vec{x}'_{\sf o}) \stackrel{{\sf def}}{=} \|\vec{x}'_{\sf o}-\vec{x}_{\sf p}\|_2^2 - (\vec{\nu}_k\cdot  \vec{x}'_{\sf o}+\beta_k) ^2 \leqslant 0, \nonumber\\
        &&\qquad  \qquad \qquad \qquad  \quad \quad  \;\;\;\;\;k=1,2,\ldots, s;
        \label{eqn:NonLinearConstraints_Problem1}
    \end{eqnarray}
    where $\|\cdot\|_2$ denotes the Euclidean 2-norm. 
\end{problem}
\hrule
\vspace{0.1in}
\vspace{0.1in}
\hrule
\begin{theorem} 
    \label{Theorem:ConfinementOfNewTrajectory}
    Given the original linear control system of \eqref{eqn:LinearSystemModel} and Assumption~\ref{assumption:ClosedFeasibleRegionOfTheReferenceState} $\sim$ \ref{assumption:PMatrixEqualEigenValue}, where the change of operational region happens at $t_{1}$. Denote $\vec{x}(t_{1})$ as $\vec{x}_{{\sf p}}$. Starting from $t_{1}$, if we only change the reference state from $\vec{x}_{{\sf o}}$ to \emph{any fixed} $\vec{x}'_{{\sf o}} \in \mathcal{R}_{{\sf o}}$, while keep other parts of the linear control system unchanged, i.e. the new linear control system becomes \eqref{eqn:NewLinearSystemModel}. Then the future trajectory of $\vec{x}(t)$ ($t \geqslant t_{1}$), denoted as $\{\vec{x}(t)\}_{t \geqslant t_{1}}$, will never exceed the hyper ellipsoid (in fact, hyper sphere, due to Assumption~\ref{assumption:PMatrixEqualEigenValue}) defined by 
    \begin{eqnarray}
    \mathcal{E}''' 
    &\stackrel{{\sf def}}{=}&
    \left\{ \vec{\xi} \, \big| (\vec{\xi} - \vec{x}'_{{\sf o}})^{{\sf T}} P (\vec{\xi} - \vec{x}'_{{\sf o}}) \leqslant (\vec{x}_{{\sf p}} - \vec{x}'_{{\sf o}})^{{\sf T}} P (\vec{x}_{{\sf p}} - \vec{x}'_{{\sf o}}), \right. \nonumber \\
    && \left. \vec{\xi} \in \mathbb{R}^{n} \right\}, \label{eqn:ConfinementOfNewTrajectory}
    \end{eqnarray}
    \noindent where $P$ (as well as $Q$) is (are) the original solution to the Lyapunov equation~\eqref{eqn:LyapunovEquation} of the original linear control system~\eqref{eqn:LinearSystemModel}.
\end{theorem}
\hrule
\vspace{0.1in}
\begin{proof}
Let us define the following function of the trajectory of $\vec{x}(t)$ ($t \geqslant t_{1}$):
\begin{eqnarray}
    v(\vec{x}(t)) \stackrel{{\sf def}}{=} (\vec{x} - \vec{x}'_{{\sf o}})^{{\sf T}} P (\vec{x} - \vec{x}'_{{\sf o}}). \label{eqn:v}
\end{eqnarray}
\noindent Then $\forall t \geqslant t_{1}$, 
\begin{eqnarray}
\dot{v} &=& \dot{\vec{x}}^{{\sf T}} P (\vec{x} - \vec{x}'_{{\sf o}}) + (\vec{x} - \vec{x}'_{{\sf o}})^{{\sf T}} P \dot{\vec{x}} \nonumber \\
&=& ((A-BK)(\vec{x} - \vec{x}'_{{\sf o}}))^{{\sf T}} P (\vec{x} - \vec{x}'_{{\sf o}}) \nonumber \\
&& + (\vec{x} - \vec{x}'_{{\sf o}})^{{\sf T}} P ((A-BK)(\vec{x} - \vec{x}'_{{\sf o}})) \quad\;\;\; \mbox{(due to \eqref{eqn:NewLinearSystemModel})} \nonumber \\
&=& (\vec{x} - \vec{x}'_{{\sf o}})^{{\sf T}} A_{{\sf cl}}^{{\sf T}} P (\vec{x} - \vec{x}'_{{\sf o}}) \nonumber \\
&& + (\vec{x} - \vec{x}'_{{\sf o}})^{{\sf T}} P A_{{\sf cl}}^{{\sf T}} (\vec{x} - \vec{x}'_{{\sf o}}) \quad\quad\quad\; \mbox{($A_{{\sf cl}} \stackrel{{\sf def}}{=} (A - BK)$)} \nonumber \\
&=& (\vec{x} - \vec{x}'_{{\sf o}})^{{\sf T}} (A_{{\sf cl}}^{{\sf T}}P + P A_{{\sf cl}}^{{\sf T}}) (\vec{x} - \vec{x}'_{{\sf o}}) \nonumber \\
&=& (\vec{x} - \vec{x}'_{{\sf o}})^{{\sf T}} (-Q) (\vec{x} - \vec{x}'_{{\sf o}}) \qquad\qquad\qquad \mbox{(due to \eqref{eqn:LyapunovEquation})} \nonumber \\
&<& 0. \qquad\qquad\qquad\qquad\quad\;\; \mbox{($Q$ is positive definite)} \nonumber
\end{eqnarray}
\noindent Therefore $\forall t \geqslant t_{1}$, 
\begin{eqnarray}
&& v(\vec{x}(t)) = (\vec{x}(t) - \vec{x}'_{{\sf o}})^{{\sf T}} P (\vec{x}(t) - \vec{x}'_{{\sf o}}) \quad \mbox{(see \eqref{eqn:v})} \nonumber \\
&\leqslant& v(\vec{x}(t_{1})) = v(\vec{x}_{{\sf p}}) \nonumber \\
&=& (\vec{x}_{{\sf p}} - \vec{x}'_{{\sf o}})^{{\sf T}} P (\vec{x}_{{\sf p}} - \vec{x}'_{{\sf o}}) \quad \mbox{(see \eqref{eqn:v})}.
\end{eqnarray}
\noindent In other words, $\forall t \geqslant t_{1}$, $\vec{x}(t) \in \mathcal{E}'''$. 
%
%
\end{proof}
\vspace{0.1in}

\vspace{0.1in}
\hrule
\begin{corollary} [Validity of ORSOP]
    \label{Corollary:ValidityOfOrsop}
    If we apply the solution to the ORSOP problem (see Problem~\ref{Problem_I}), denoted as $\vec{x}'^{*}_{{\sf o}}$, to the new linear control system \eqref{eqn:NewLinearSystemModel}, then requirement {\bf R1} $\sim$ {\bf R3} are all satisfied. Particularly, the Lyapunov ellipsoid $\mathcal{E}''$ requested by requirement {\bf R1} is given by 
    \begin{eqnarray}
    && \mathcal{E}''^{*} = E(\vec{x}_{{\sf p}}, \vec{x}'^{*}_{{\sf o}}, P) \nonumber \\
    &=& \left\{ \vec{\xi} \, \big| (\vec{\xi} - \vec{x}'^{*}_{{\sf o}})^{{\sf T}} P (\vec{\xi} - \vec{x}'^{*}_{{\sf o}}) 
    \leqslant 
    (\vec{x}_{{\sf p}} - \vec{x}'^{*}_{{\sf o}})^{{\sf T}} P (\vec{x}_{{\sf p}} - \vec{x}'^{*}_{{\sf o}}), \right. \nonumber \\
    && \left. \vec{\xi} \in \mathbb{R}^n \right\}. 
    \label{eqn:OrsopNewLyapunovEllipsoid}
    \end{eqnarray}
\end{corollary}
\hrule
\vspace{0.1in}
\begin{proof}

Due to Assumption~\ref{assumption:PMatrixEqualEigenValue}, $\mathcal{E}''^{*}$ is a hyper sphere centered at $\vec{x}'^{*}_{\sf o}$, and has a radius of $||\vec{x}'^{*}_{\sf o} - \vec{x}_{\sf p}||_2$. Meanwhile, according to analytical geometry, the distance between $\vec{x}'^{*}_{\sf o}$ to hyper plane $\vec{v}_k \cdot \vec{x} + \beta_{k} = 0$ (i.e. the boundary of the new linear operational constraint $\vec{v}_{k} \cdot \vec{x} + \beta_{k} \leqslant 0$) is $\sqrt{(\vec{v}_{k} \cdot \vec{x}'^{*}_{{\sf o}} + \beta_{k})^2}$. Combined with Assumption~\ref{assumption:Rescueable}, \eqref{eqn:NonLinearConstraints_Problem1} implies $\mathcal{E}''^{*} \cap \mathcal{F}' = \varnothing$.

Meanwhile, if we choose $\vec{x}'_{{\sf o}} = \vec{x}'^{*}_{{\sf o}}$ for the new linear control system \eqref{eqn:NewLinearSystemModel} from $t_{1}$, Theorem~\ref{Theorem:ConfinementOfNewTrajectory} implies $\{\vec{x}(t)\}_{t \geqslant t_{1}} \subseteq \mathcal{E}''^{*}$. 

Thirdly, comparing \eqref{eqn:NewLyapunovEllipsoid} and \eqref{eqn:OrsopNewLyapunovEllipsoid}, we see $\mathcal{E}''^{*}$ is the requested $\mathcal{E}''$.

In summary, requirement {\bf R1} is satisfied.

Meanwhile, \eqref{eqn:LinearConstraints_Problem1} means requirement {\bf R2} is satisfied.

Thirdly, due to Assumption~\ref{assumption:PMatrixEqualEigenValue}, the objective function \eqref{eqn:ObjectiveFunctionReachability_Problem1} means requirement {\bf R3} is satisfied. 

%
\end{proof}
\vspace{0.1in}

\section{Proposed Solution}
\label{sect:ProposedSolution}

In this section, we propose our solution to Problem~\ref{Problem_I} (aka the ORSOP).

To meet the real-time demand, ideally, we want the solution to be analytical. 

Before we proceed, note constraint \eqref{eqn:LinearConstraints_Problem1} of Problem~\ref{Problem_I}, defines the \emph{compact} (i.e. closed and bounded) feasible region of the reference state $\mathcal{R}_{\sf o}$. For ease of narration, let us denote the boundary of $\mathcal{R}_{\sf o}$ as $\partial \mathcal{R}_{\sf o}$. Note as $\mathcal{R}_{\sf o}$ is compact, $\partial \mathcal{R}_{\sf o} \subseteq \mathcal{R}_{\sf o}$.

Meanwhile, the objective function \eqref{eqn:ObjectiveFunctionReachability_Problem1} of Problem~\ref{Problem_I} implies that the solution $\vec{x}'^*_{\sf o}$ is affected by the present plant state $\vec{x}_{\sf p}$. Therefore, we can analyze $\vec{x}'^*_{\sf o}$ case by case depending on $\vec{x}_{\sf p}$.
\begin{description}
    \item[\textbf{Case 1}:] 
    $\vec{x}_{\sf p} \in \mathcal{R}_{\sf o}$.
    \item[\textbf{Case 2}:] 
    $\vec{x}_{\sf p} \notin \mathcal{R}_{\sf o}$.
\end{description}

\subsection{Optimal Solution for Case 1}
\label{subsect:OptimalSolutionForCase1}

Case~1 is trivial. The solution is analytical and is $\vec{x}'^*_{\sf o}=\vec{x}_{\sf p}$, as this sets the objective function \eqref{eqn:ObjectiveFunctionReachability_Problem1} to the global minimum: $f(\vec{x}'^*_{\sf o})=\|\vec{x}'^*_{\sf o}-\vec{x}_{\sf p}\|_2^2=0$. Meanwhile, the solution $\vec{x}'^*_{\sf o}=\vec{x}_{\sf p}$ satisfies both constraint \eqref{eqn:LinearConstraints_Problem1} and \eqref{eqn:NonLinearConstraints_Problem1}. Specifically, 
\begin{enumerate}
    \item as $\vec{x}_{\sf p} \in \mathcal{R}_{\sf o}$, and $\mathcal{R}_{\sf o}$ is defined by \eqref{eqn:LinearConstraints_Problem1}, $\vec{x}'^{*}_{{\sf o}}$ ($= \vec{x}_{\sf p}$) hence complies with \eqref{eqn:LinearConstraints_Problem1}.
    \item When $\vec{x}'^*_{\sf o}=\vec{x}_{\sf p}$, 
    $q_k(\vec{x}'^*_{\sf o}) =-(\vec{\nu}_k\cdot \vec{x}'^*_{\sf o}+\beta_k)^2\leqslant 0$ ($k=1,2,\cdots, s$), i.e. \eqref{eqn:NonLinearConstraints_Problem1} sustains.
\end{enumerate}
%

\subsection{Optimal Solution for Case 2}
\label{subsect:OptimalSolutionForCase2}

Because $\vec{x}'^*_{\sf o}$ must reside in\footnote{For clarification, in this paper, a plant state is said to be \enquote{inside} a region when the plant state resides in the interior of the region excluding the region boundaries. On the other hand, a plant state is said to be \enquote{in} the region when the plant state resides in the interior or on the boundaries.} the feasible region of the reference state $\mathcal{R}_{\sf o}$, we cannot assign $\vec{x}'^*_{\sf o} = \vec{x}_{\sf p}$ since $\vec{x}_{\sf p} \notin \mathcal{R}_{\sf o}$ in Case~2.

In addition, it is trivial that $\vec{x}'^*_{\sf o}$ cannot exist inside $\mathcal{R}_{\sf o}$, because we can always find another reference state (denoted as $\vec{x}'^{**}_{\sf o}$) that is along the direction from $\vec{x}'^*_{\sf o}$ to $\vec{x}_{\sf p}$ and on the boundary of $\mathcal{R}_{\sf o}$ (i.e., $\vec{x}'^{**}_{\sf o} \in \partial \mathcal{R}_{\sf o}$), so that $\|\vec{x}'^{**}_{\sf o}-\vec{x}_{\sf p}\|_2 < \|\vec{x}'^{*}_{\sf o}-\vec{x}_{\sf p}\|_2$. 
\vspace{0.1in}
\hrule
\begin{remark}
    \label{remark:SolutionMustOnBoundary_Case3}
    In Case~2, the optimal solution $\vec{x}'^{*}_{{\sf o}}$, if exists, must reside on some boundaries of $\mathcal{R}_{\sf o}$ (i.e., $\vec{x}'^*_{\sf o}\in \partial \mathcal{R}_{\sf o}$).
\end{remark}
\hrule
\vspace{0.1in}
To find the optimal solution $\vec{x}'^{*}_{{\sf o}}$ for Case~2, we notice the nonlinear constraint \eqref{eqn:NonLinearConstraints_Problem1} complicates our analysis. To simply, we propose the following \emph{3-step procedure}.
%
\begin{enumerate}
    \item \textit{Step 1}: Simplify Problem~\ref{Problem_I} to the below Problem~\ref{Problem_II} by removing the nonlinear constraint \eqref{eqn:NonLinearConstraints_Problem1}. Problem~\ref{Problem_II} is a classical optimization problem that can be analytically solved via the KKT method (see Section~\ref{subsect:KktOptimization}). Denote the thus derived analytical optimal solution to Problem~\ref{Problem_II} as $\vec{x}'^{\star}_{\sf o}$.
    \vspace{0.1in}
    \hrule
    \begin{problem}
        \label{Problem_II}  
        \begin{eqnarray}
            &&{\sf min}_{\vec{x}'_{\sf o}}    \; \;  \left( f(\vec{x}'_{\sf o})= \|\vec{x}'_{\sf o}-\vec{x}_{\sf p}\|_2^2 \right), \label{eqn:SimplifiedObjectiveFunction_Problem2}\\
            \mbox{s.t.} \;\;  && g_j(\vec{x}'_{\sf o})=\vec{\omega}_j\cdot  \vec{x}'_{\sf o}+b_j \leqslant 0,\;\; j=1,2,\ldots, r. \qquad \quad \label{eqn:SimplifiedLinearConstraint_Problem2}
        \end{eqnarray}
    \end{problem}
    \hrule
    \vspace{0.1in}
    \item \textit{Step 2}: Check if the $\vec{x}'^{\star}_{\sf o}$ from \textit{Step 1} complies with the nonlinear constraint \eqref{eqn:NonLinearConstraints_Problem1} of Problem~\ref{Problem_I}. If so, return $\vec{x}'^{\star}_{\sf o}$ for Problem~\ref{Problem_II} as the optimal solution $\vec{x}'^*_{\sf o}$ for Problem~\ref{Problem_I}. Otherwise, proceed to \textit{Step 3}.
    \item \textit{Step 3}: Apply the IPM-based Newton's method (see Section~\ref{subsect:NewtonsMethod}) to numerically search for the solution for Problem~\ref{Problem_I}. If the search finds a solution $\vec{x}'^*_{\sf o}$, return this $\vec{x}'^*_{\sf o}$ as the optimal solution for Problem~\ref{Problem_I}. Return ``failure'' otherwise.
\end{enumerate}

Note, due to \textit{Step 3}, the above \emph{3-step procedure} uses the well-known IPM-based Newton's method as its fall-back plan in the search for $\vec{x}'^{*}_{{\sf o}}$. Therefore, we have the following trivial proposition.
\vspace{0.1in}
\hrule
\begin{proposition}
If Problem~\ref{Problem_I}'s optimal solution of $\vec{x}'^{*}_{{\sf o}}$ for Case~2 can be found by the IPM-based Newton's method alone, then $\vec{x}'^{*}_{{\sf o}}$ can be found by the proposed 3-step procedure for Case~2.
\end{proposition}
\hrule
\vspace{0.1in}

However, there are still two details of the 3-step procedure that need further clarification: how to conduct the ``KKT method'' in Step 1, and how to conduct the ``IPM-based Newton's method'' in Step 3. These will be elaborated in the following respectively by Section~\ref{subsect:Step1OfThe3StepProcedure} and \ref{subsect:Step3OfThe3StepProcedure}.

\subsection{Step 1 of the 3-Step Procedure}
\label{subsect:Step1OfThe3StepProcedure}

In this sub-section, we shall elaborate the ``KKT method'' in \textit{Step 1} of the proposed 3-step procedure in Section~\ref{subsect:OptimalSolutionForCase2}.

In \textit{Step 1}, the Lagrange function for Problem~\ref{Problem_II} is given as following (see Lemma~\ref{lemma:KKT}): 
\begin{eqnarray}
    \label{eqn:LagrangeFunction_case2}
    &&L(\vec{x}'_{\sf o})= f(\vec{x}'_{\sf o})+\sum_{j=1}^{r} \mu_jg_j(\vec{x}'_{\sf o}).
\end{eqnarray}
The partial derivatives (w.r.t $\vec{x}'_{\sf o}$) of the involved functions in \eqref{eqn:LagrangeFunction_case2} are given as follows:
\begin{eqnarray}
    \label{eqn:TimeDerivativeObjectiveFunction_case2}
    \frac{\partial f(\vec{x}'_{\sf o})}{\partial \vec{x}'_{\sf o}}&=&2(\vec{x}'_{\sf o}-\vec{x}_{\sf p}); \;\in \mathbb{R}^n\\
    \frac{\partial g_j(\vec{x}'_{\sf o})}{\partial \vec{x}'_{\sf o}}&=&\vec{\omega}_j \;\in \mathbb{R}^n, \;\; j=1,2,\ldots,r.
    \label{eqn:TimeDerivativeLinearConstraints_case2}
\end{eqnarray}
%
%

%
%
Based on Remark~\ref{remark:SolutionMustOnBoundary_Case3}, we can assume that the optimal solution $\vec{x}'^{\star}_{\sf o}$ resides on the intersection of exactly $l$ ($l \in \{1$, $\ldots$, $r\}$) boundaries, i.e. 
%
%
\begin{eqnarray}
\vec{x}'^{\star}_{\sf o} \in \left\{ \xi \, \big| \, \mbox{$g_{j}(\xi) = 0$ ($\forall j \in \{[1], [2], \ldots, [l]\}$) and } \right. \qquad \qquad \nonumber \\
\left. \mbox{$g_{j}(\xi) < 0$ ($\forall j \in \{1, 2, \ldots, r\} - \{[1], [2], \ldots, [l]\}$)} \right\}, \label{eqn:l-CombinationOfBoundaries}
\end{eqnarray} 
\noindent where $\forall i < j$ ($i, j \in \{1, \ldots, l\}$), we have $[i] < [j]$ and $[i], [j] \in \{1, \ldots, r\}$. Furthermore, assume $\left\{[1], [2], \ldots, [l]\right\}$ is the $\ell$th ($\ell \in \left\{1, \ldots, {r \choose l}\right\}$) distinct combination of $l$ indices from the index set $\{1, 2, \ldots, r\}$.

    %
    %
    %
%
    %
    %
   %

Let $\vec{d}_{l, \ell}$ and $W_{l, \ell}$ represent the following vector and matrix:
%
\begin{eqnarray}
    \label{eqn:vectorCurrentState2FeasibleRegion}
    \vec{d}_{l, \ell} \stackrel{{\sf def}}{=} \begin{bmatrix}
    \vec{\omega}_{[1]} \cdot \vec{x}_{\sf p}+b_{[1]}\\
    \vec{\omega}_{[2]} \cdot \vec{x}_{\sf p}+b_{[2]}\\
    \vdots\\
    \vec{\omega}_{[l]} \cdot \vec{x}_{\sf p}+b_{[l]}\\
    \end{bmatrix} \in \mathbb{R}^l,\;\;  
\end{eqnarray}
and
\begin{eqnarray}
    \label{eqn:MatrixNormVectorultiplication}
    W_{l, \ell} \stackrel{{\sf def}}{=} \begin{bmatrix}
    \vec{\omega}_{[1]} \cdot \vec{\omega}_{[1]} & \vec{\omega}_{[1]} \cdot \vec{\omega}_{[2]} &\cdots& \vec{\omega}_{[1]} \cdot \vec{\omega}_{[l]} \\
    \vec{\omega}_{[2]} \cdot \vec{\omega}_{[1]} & \vec{\omega}_{[2]} \cdot \vec{\omega}_{[2]} &\cdots& \vec{\omega}_{[2]} \cdot \vec{\omega}_{[l]}\\
    \vdots & \vdots & \cdots& \vdots\\
    \vec{\omega}_{[l]} \cdot \vec{\omega}_{[1]} & \vec{\omega}_{[l]} \cdot \vec{\omega}_{[2]}& \cdots& \vec{\omega}_{[l]} \cdot \vec{\omega}_{[l]}\\
    \end{bmatrix} \in \mathbb{R}^{l \times l}. 
\end{eqnarray}
Then, we have the following important theorem for finding the candidate solutions in Case~2:
\vspace{0.1in}
\hrule
\begin{theorem} 
    \label{Theorem:optimalSolution4Problem2_case2}
    Given $\vec{x}_{\sf p} \notin \mathcal{R}_{\sf o}$, if  \eqref{eqn:l-CombinationOfBoundaries} sustains and $W_{l, \ell}$ is invertible, then the candidate optimal solution to Problem~\ref{Problem_II} is 
    \begin{eqnarray}
    \vec{x}'^{\star}_{{\sf o}} = \vec{x}_{{\sf p}} - \frac{1}{2} \sum_{j=1}^{l} \mu^{\star}_{[j]} \vec{\omega}_{[j]}, \label{eqn:optimalSolution4Problem2InTheorem_case2}
    \end{eqnarray}
    \noindent where 
    \begin{eqnarray}
    \vec{\mu}_{l, \ell}^{\star} \stackrel{{\sf def}}{=} (\mu_{[1]}^{\star}, \mu_{[2]}^{\star}, \ldots, \mu_{[l]}^{\star})^{\sf T} = 2 W^{-1}_{l, \ell} \vec{d}_{l, \ell}. \label{eqn:vec_mu_star_l_ell}
    \end{eqnarray}
\end{theorem}
\hrule
\vspace{0.1in}
\begin{proof}

If \eqref{eqn:l-CombinationOfBoundaries} sustains, the KKT conditions listed in Lemma~\ref{lemma:KKT} shall manifest in the following form.

First, due to the complementary slackness, $\forall j \in \{1, 2, \ldots, r\} - \{[1], [2], \ldots, [l]\}$, we have $\mu^{\star}_j = 0$. $\hfill(\dagger)$

Second, due to the stationarity, the Lagrange function of \eqref{eqn:LagrangeFunction_case2} should satisfy 
\begin{eqnarray}
    && \frac{\partial L(\vec{x}'^{\star}_{\sf o})}{\partial \vec{x}'_{\sf o}} = \frac{\partial f(\vec{x}'^{\star}_{\sf o})}{\partial \vec{x}'_{\sf o}}+\sum_{j=1}^{r} \mu^{\star}_{j} \frac{\partial g_{j}(\vec{x}'^{\star}_{\sf o})}{\partial \vec{x}'_{\sf o}} \nonumber \\
    &=& \frac{\partial f(\vec{x}'^{\star}_{\sf o})}{\partial \vec{x}'_{\sf o}}+\sum_{j=1}^{l} \mu^{\star}_{[j]} \frac{\partial g_{[j]}(\vec{x}'^{\star}_{\sf o})}{\partial \vec{x}'_{\sf o}} \;\; \mbox{ (due to $(\dagger)$)} \nonumber \\
    &=& 2(\vec{x}'^{\star}_{\sf o}-\vec{x}_{\sf p})+\sum_{j=1}^{l} \mu^{\star}_{[j]} \vec{\omega}_{[j]} \;\; \mbox{ (due to \eqref{eqn:TimeDerivativeObjectiveFunction_case2} and \eqref{eqn:TimeDerivativeLinearConstraints_case2})} \nonumber \\
    &=& \mathbf{0} \;\; \mbox{ (due to the stationarity of the KKT conditions)}.\nonumber \\
    && \label{eqn:DerivativeOfLagrangeFunction_Case2}
\end{eqnarray}
\noindent Then, from \eqref{eqn:DerivativeOfLagrangeFunction_Case2}, we have the following solution:  
\begin{eqnarray}
    \vec{x}'^{\star}_{{\sf o}}
    = \vec{x}_{\sf p}-\frac{1}{2}\sum_{j=1}^{l} \mu_{[j]}^{\star} \vec{\omega}_{[j]}. \label{eqn:OptimalSolution_x_l_Case2}
\end{eqnarray}
However, the Lagrange multipliers, i.e. $\vec{\mu}^{\star}_{l, \ell} \stackrel{{\sf def}}{=} (\mu_{[1]}^{\star},\mu_{[2]}^{\star},\cdots,\mu_{[l]}^{\star} )^{\sf T}$, are still unknown. This can be solved by the following conditions included in \eqref{eqn:l-CombinationOfBoundaries}:
\begin{eqnarray} 
    & g_{[1]}(\vec{x}'^{\star}_{\sf o})= \vec{\omega}_{[1]} \cdot \vec{x}'^{\star}_{\sf o}+b_{[1]} = 0; \nonumber \\
    & g_{[2]}(\vec{x}'^{\star}_{\sf o})= \vec{\omega}_{[2]} \cdot \vec{x}'^{\star}_{\sf o}+b_{[2]} =0; \nonumber \\
    & \qquad \qquad \vdots \nonumber\\
    & g_{[l]}(\vec{x}'^{\star}_{\sf o})=\vec{\omega}_{[l]} \cdot \vec{x}'^{\star}_{\sf o}+b_{[l]} =0. \label{eq_slackness_j}
\end{eqnarray}
Substituting \eqref{eqn:OptimalSolution_x_l_Case2} into \eqref{eq_slackness_j} leads to the following set of equations: 
\begin{eqnarray}
    \label{eqn:}
    \vec{\omega}_{[1]} \cdot \vec{x}_{\sf p}-\frac{1}{2}( \mu^{\star}_{[1]} \vec{\omega}_{[1]} \cdot \vec{\omega}_{[1]}+\mu^{\star}_{[2]} \vec{\omega}_{[1]} \cdot \vec{\omega}_{[2]}+\cdots +\mu^{\star}_{[l]} \vec{\omega}_{[1]} \cdot \vec{\omega}_{[l]}) \nonumber \\
    +b_{[1]}=0; \nonumber\\
    \vec{\omega}_{[2]} \cdot \vec{x}_{\sf p}-\frac{1}{2} (\mu^{\star}_{[1]} \vec{\omega}_{[2]} \cdot \vec{\omega}_{[1]} +\mu^{\star}_{[2]} \vec{\omega}_{[2]} \cdot \vec{\omega}_{[2]}+\cdots +\mu^{\star}_{[l]} \vec{\omega}_{[2]} \cdot \vec{\omega}_{[l]}) \nonumber \\
    +b_{[2]}=0;  \nonumber\\    
    \vdots \;\;\;\;\;\;\;\;\;\;\;\;\;\;  \;\;\;\;\;\;\;\;\;\;\;\;\;\; \;\;\;\;\;\;\;\;\;\;\;\;\;\; \nonumber\\
    \vec{\omega}_{[l]} \cdot \vec{x}_{\sf p}-\frac{1}{2} (\mu^{\star}_{[1]} \vec{\omega}_{[l]} \cdot \vec{\omega}_{[1]} +\mu^{\star}_{[2]} \vec{\omega}_{[l]} \cdot \vec{\omega}_{[2]}+\cdots+\mu^{\star}_{[l]} \vec{\omega}_{[l]} \cdot \vec{\omega}_{[l]}) \nonumber \\
    +b_{[l]}=0, \nonumber
\end{eqnarray}
which can be concatenated as
\begin{eqnarray}
    \label{eqn:LagrangeMultiplierSolutionIntermidiateMatrixForm_case2}
    &&
    \begin{bmatrix}
        \vec{\omega}_{[1]} \cdot \vec{\omega}_{[1]} & \vec{\omega}_{[1]} \cdot \vec{\omega}_{[2]} & \cdots &\vec{\omega}_{[1]} \cdot \vec{\omega}_{[l]}\\
        \vec{\omega}_{[2]} \cdot \vec{\omega}_{[1]} & \vec{\omega}_{[2]} \cdot \vec{\omega}_{[2]} & \cdots &\vec{\omega}_{[2]} \cdot \vec{\omega}_{[l]}\\
        \vdots & \vdots & \cdots &\vdots\\
        \vec{\omega}_{[l]} \cdot \vec{\omega}_{[1]} & \vec{\omega}_{[l]} \cdot \vec{\omega}_{[2]} & \cdots &\vec{\omega}_{[l]} \cdot \vec{\omega}_{[l]}\\
    \end{bmatrix} 
    \begin{bmatrix}
        \mu^{\star}_{[1]}\\
        \mu^{\star}_{[2]}\\
        \vdots \\
        \mu^{\star}_{[l]}\\
    \end{bmatrix} \nonumber \\
    &=& 2
    \begin{bmatrix}
        \vec{\omega}_{[1]} \cdot \vec{x}_{\sf p}+b_{[1]}\\
        \vec{\omega}_{[2]} \cdot \vec{x}_{\sf p}+b_{[2]}\\
        \vdots\\
        \vec{\omega}_{[l]} \cdot \vec{x}_{\sf p}+b_{[l]}\\
    \end{bmatrix}.
\end{eqnarray}
Using \eqref{eqn:vectorCurrentState2FeasibleRegion} and \eqref{eqn:MatrixNormVectorultiplication}, \eqref{eqn:LagrangeMultiplierSolutionIntermidiateMatrixForm_case2} can be expressed by 
\begin{eqnarray}
    %
    %
    W_{l, \ell} \vec{\mu}^{\star}_{l, \ell} = 2\vec{d}_{l, \ell}. \nonumber
\end{eqnarray}
As $W_{l, \ell}$ is invertible, we have
\begin{eqnarray}
    %
    %
    \vec{\mu}^{\star}_{l, \ell} = 2W^{-1}_{l, \ell} \vec{d}_{l, \ell}. \nonumber
\end{eqnarray}

This concludes the proof. 
%
%
\end{proof}
\vspace{0.1in}

Based on Theorem~\ref{Theorem:optimalSolution4Problem2_case2}, the KKT method used in \textit{Step 1} can be formally defined by Algorithm~\ref{alg:KktMethodUsedInStep1}.

\begin{algorithm}[htbp]
\caption{\textbf{KKT method used in \textit{Step 1}}}
\label{alg:KktMethodUsedInStep1}
\setcounter{ctr}{0}
\begin{tabbing}
888\=88\=88\=88\=88\=88\=88\=88\=88\=88\=88\kill
\textbf{function} KktMethodUsedInStep1 ( \\
\> \textbf{input}: Problem~\ref{Problem_II}; \\
\> \textbf{output}: $\vec{x}'^{\star}_{{\sf o}}$, i.e. the solution to Problem~\ref{Problem_II} \\
):  \\
\stp{} \> Set of candidate solutions $\mathcal{X} := \varnothing$; \\
\stp{} \> \textbf{for} $l$ in $\{1, \ldots, r\}$ \textbf{do} \\
\stp{} \>\> \textbf{for} $\ell$ in $\left\{1, \ldots, {r \choose l} \right\}$ \textbf{do} \\
\stp{} \>\>\> Create the $\ell$th distinct combination of $l$ indices  \\
\>\>\> from the index set $\{1, \ldots, r\}$, and denote this \\
\>\>\> indices combination as set $\left\{[1],  \ldots, [l]\right\}$; \\
\stp{} \>\>\> $\vec{x}'^{\star}_{{\sf o}, l, \ell} := $ NaN;  $\,$ //NaN: Not a Number\\
\stp{} \>\>\>  \textbf{if} $W_{l, \ell}$ is invertible \textbf{then} \\
\stp{} \>\>\>\> $\vec{\mu}^{\star}_{l, \ell} := 2 W^{-1}_{l, \ell}\vec{d}_{l, \ell}$; $\,$ //see \eqref{eqn:vec_mu_star_l_ell} of Theorem~\ref{Theorem:optimalSolution4Problem2_case2} \\
\stp{} \>\>\>\> $\vec{x}'^{\star}_{{\sf o}, l, \ell} := \vec{x}_{{\sf p}} - \frac{1}{2} \sum_{j=1}^{l} \mu^{\star}_{[j]} \vec{\omega}_{[j]}$; \\
\>\>\>\> //see \eqref{eqn:optimalSolution4Problem2InTheorem_case2} of Theorem~\ref{Theorem:optimalSolution4Problem2_case2} \\
\stp{} \>\>\>\> //check if $\vec{x}'^{\star}_{{\sf o},l,\ell}$ satisfies presumption \eqref{eqn:l-CombinationOfBoundaries}: \\
\stp{} \>\>\>\> \textbf{if} $\exists [j] \in \{[1], \ldots, [l]\}$ s.t. $g_{[j]}(\vec{x}'^{\star}_{{\sf o},l,\ell}) \neq 0$ \\
\>\>\>\> \textbf{then} $\vec{x}'^{\star}_{{\sf o},l,\ell} := \mbox{NaN}$; \textbf{endif}; \\
\stp{} \>\>\>\> \textbf{if} $\exists j \in \{1, \ldots, r\} -\{[1], \ldots, [l]\}$ s.t. $g_{j}(\vec{x}'^{\star}_{{\sf o},l,\ell}) \geqslant 0$ \\
\>\>\>\> \textbf{then} $\vec{x}'^{\star}_{{\sf o},l,\ell} := \mbox{NaN}$; \textbf{endif};\\
\stp{} \>\>\> \textbf{endif}; \\
\stp{} \>\>\> \textbf{if} $\vec{x}'^{\star}_{{\sf o},l,\ell} \neq \mbox{NaN}$ \textbf{then} $\mathcal{X} := \mathcal{X} \cup \{\vec{x}'^{\star}_{{\sf o},l,\ell}\}$; \textbf{endif}; \\
\stp{} \>\> \textbf{endfor}; \\
\stp{} \> \textbf{endfor}; \\
\stp{} \> \textbf{if} $\mathcal{X} \neq \varnothing$ \textbf{then} \\
\stp{} \>\> enumerate all the elements in $\mathcal{X}$ to find $\vec{x}'^{\star}_{{\sf o}} \in \mathcal{X}$ that \\
\>\> minimizes $f(\vec{x}'_{{\sf o}}) \stackrel{{\sf def}}{=} (|| \vec{x}'_{{\sf o}} - \vec{x}_{{\sf p}} ||_2)^2$; \\
\>\> //see \eqref{eqn:SimplifiedObjectiveFunction_Problem2} of Problem~\ref{Problem_II}\\
\stp{} \>\> \textbf{return} $\vec{x}'^{\star}_{{\sf o}}$; \\
\stp{} \> \textbf{else} \textbf{return} NaN; \textbf{endif};
\end{tabbing}
\end{algorithm}


\subsection{Step 3 of the 3-Step Procedure}
\label{subsect:Step3OfThe3StepProcedure}

In this sub-section, we shall elaborate the ``IPM-based Newton's method'' in \textit{Step 3} of the proposed 3-step procedure in Section~\ref{subsect:OptimalSolutionForCase2}.

Based on Section~\ref{subsect:NewtonsMethod}, we adopt the natural logarithmic approximation form of the indicator function (see \eqref{eqn:IndicationFunctionApproximate_Log}), so as to re-write Problem~\ref{Problem_I} into an unconstrained form (see \eqref{eqn:UnconstrainedObjectiveFunction4NewtonMethod}): 
\begin{eqnarray}
    & {\sf min}_{\vec{x}'_{{\sf o}}} \left( F(\vec{x}'_{\sf o}) \right), \nonumber \\
    & \mbox{where } F(\vec{x}'_{\sf o}) = f(\vec{x}'_{\sf o})-\frac{1}{\lambda} \sum_{j=1}^{r} {\sf ln} \big(-g_j(\vec{x}'_{\sf o})\big) \nonumber \\
    & \hspace{1.3cm} -\frac{1}{\lambda} \sum_{k=1}^{s} {\sf ln} \big(-q_k(\vec{x}'_{\sf o})\big).
    \label{eqn:Step3UnconstrainedForm}
\end{eqnarray}
The Newton's method needs the gradient $\nabla F(\vec{x}'_{\sf o})$ and Hessian matrix $\nabla	^2F(\vec{x}'_{\sf o})$ of $F(\vec{x}'_{\sf o})$. They are derived as follows. 

\subsubsection{Gradient}
\begin{eqnarray}
    \nabla	F(\vec{x}'_{\sf o})=\frac{\partial F(\vec{x}'_{\sf o})}{\partial \vec{x}'_{\sf o}}
    &=&2(\vec{x}'_{\sf o}-\vec{x}_{\sf p})- \frac{1}{\lambda}  \sum_{j=1}^{r} (g_j(\vec{x}'_{\sf o}))^{-1} \frac{\partial g_j(\vec{x}'_{\sf o})}{\partial \vec{x}'_{\sf o}}  \nonumber\\
    &&- \frac{1}{\lambda}  \sum_{k=1}^{s} (q_k(\vec{x}'_{\sf o}))^{-1} \frac{\partial q_k(\vec{x}'_{\sf o})}{\partial \vec{x}'_{\sf o}}, 
    \label{eqn:GradientFunction_Fx_NewtonsMethod}
\end{eqnarray}
where 
\begin{eqnarray}
    \frac{\partial g_j(\vec{x}'_{\sf o})}{\partial \vec{x}'_{\sf o}}=\frac{\partial (\vec{\omega}_j\cdot \vec{x}'_{\sf o}+b_j)}{\partial \vec{x}'_{\sf o}}=\vec{\omega}_j \;\in \mathbb{R}^n, \label{eqn:GradientFunction_gj}
\end{eqnarray}
and 
\begin{eqnarray}
    \frac{\partial q_k(\vec{x}'_{\sf o})}{\partial \vec{x}'_{\sf o}}&=&\frac{\partial \Big((\vec{x}'_{\sf o}-\vec{x}_{\sf p})^2-(\vec{\nu}_k\cdot \vec{x}'_{\sf o}+\beta _k)^2\Big)}{\partial \vec{x}'_{\sf o}} \nonumber \\
    &=&2(\vec{x}'_{\sf o}-\vec{x}_{\sf p})-2(\vec{\nu}_k \cdot \vec{x}'_{\sf o}+\beta _k)\vec{\nu}_k\nonumber\\
    &=&2(\vec{x}'_{\sf o}-\vec{x}_{\sf p})-2\vec{\nu}_k\vec{\nu}_k^T \vec{x}'_{\sf o}-2\beta _k\vec{\nu}_k \;\in \mathbb{R}^n. \label{eqn:GradientFunction_qk}
\end{eqnarray}
%

\subsubsection{Hessian matrix} 

To calculate the Hessian matrix $\nabla	^2F(\vec{x}'_{\sf o})$, we first use \eqref{eqn:GradientFunction_gj} and \eqref{eqn:GradientFunction_qk}, to respectively derive 
\begin{eqnarray}
    \label{eqn:HessianMatrix_gj}
    \frac{\partial^2 g_j(\vec{x}'_{\sf o})}{\partial \vec{x}'^2_{\sf o}} &=& \mathbf{0}_n, \\
    \label{eqn:HessianMatrix_qk}
    \frac{\partial^2 q_k(\vec{x}'_{\sf o})}{\partial \vec{x}'^2_{\sf o}} &=& 2I_n- 2\vec{\nu}_k\vec{\nu}_k^{\sf T} \;\in \mathbb{R}^{n\times n},
\end{eqnarray}
\noindent where $I_n \in \mathbb{R}^{n\times n}$ is the identity matrix, and $\mathbf{0}_n \in \mathbb{R}^{n\times n}$ is the zero matrix. With \eqref{eqn:HessianMatrix_gj} and \eqref{eqn:HessianMatrix_qk}, we have 
\begin{eqnarray}
    & \nabla^2 F(\vec{x}'_{\sf o})=\frac{\partial^2 F(\vec{x}'_{\sf o})}{\partial \vec{x}'^2_{\sf o}}
    \hspace{5cm} \nonumber \\
    & = 2I_n -\frac{1}{\lambda}  \sum_{j=1}^{r} \Bigg[-(g_j(\vec{x}'_{\sf o}))^{-2}\Big(\frac{\partial g_j(\vec{x}'_{\sf o})}{\partial \vec{x}'_{\sf o}}\Big)^2+(g_j(\vec{x}'_{\sf o}))^{-1}\frac{\partial^2 g_j(\vec{x}'_{\sf o})}{\partial \vec{x}'^2_{\sf o}} \Bigg] \nonumber \\
    & - \frac{1}{\lambda}  \sum_{k=1}^{s} \Bigg[-(q_k(\vec{x}'_{\sf o}))^{-2}\Big(\frac{\partial q_k(\vec{x}'_{\sf o})}{\partial \vec{x}'_{\sf o}}\Big)^2+(q_k(\vec{x}'_{\sf o}))^{-1}\frac{\partial^2 q_k(\vec{x}'_{\sf o})}{\partial \vec{x}'^2_{\sf o}} \Bigg].
    \label{eqn:HessianMatrix_Fx_NewtonsMethod}
\end{eqnarray}

With \eqref{eqn:GradientFunction_Fx_NewtonsMethod} and \eqref{eqn:HessianMatrix_Fx_NewtonsMethod}, the optimal solution for \eqref{eqn:Step3UnconstrainedForm} can be searched iteratively by (see \eqref{eqn:NewtonIterationFunction}):
\begin{eqnarray}
    \label{eqn:IterationFunction_NewtonsMethod}
    \vec{x}'^{(\imath+1)}_{\sf o}=\vec{x}'^{(\imath)}_{\sf o}-\eta [\nabla ^2F(\vec{x}'^{(\imath)}_{\sf o})]^{-1} \nabla F(\vec{x}'^{(\imath)}_{\sf o}),
\end{eqnarray}
where $\eta$ is the iteration step size (which is fixed in this paper). The iteration ending conditions/operations are described by {\bf E1} and {\bf E2} in Section~\ref{subsect:NewtonsMethod}, which will not be repeated here.

Note, there is one more implementation detail to take care of. As required by $(*)$ in Section~\ref{subsect:NewtonsMethod}, throughout the iterations $\imath = 0, 1, \ldots $, we need to assert
\begin{eqnarray}
    \forall j \in \{1, \ldots, r\}, \quad g_j(\vec{x}'^{(\imath)}_{\sf o}) < 0; \label{eqn:IPMPerStepAssert_g_j} \\
    \mbox{and } \forall k \in \{1, \ldots, s\}, \quad q_k(\vec{x}'^{(\imath)}_{\sf o}) < 0. \label{eqn:IPMPerStepAssert_q_k}
\end{eqnarray} 
\noindent Otherwise, we need to stop the iteration and claim the failure of the IPM-based Newton's method.

The above also implies that the choice of $\vec{x}'^{(0)}_{\sf o}$ must satisfy \eqref{eqn:IPMPerStepAssert_g_j} and \eqref{eqn:IPMPerStepAssert_q_k}. Otherwise, we need to claim failure at the start of the IPM-based Newton's method. How to best choose $\vec{x}'^{(0)}_{\sf o}$ remains as an open problem. In this paper, we propose a naive solution: simply choose $\vec{x}'^{(0)}_{\sf o} = \vec{x}_{\sf o}$, i.e. the original reference state. Note this naive solution only makes the evaluation comparisons more pessimistic on our proposed solution.

\subsection{Relaxation of Assumption~\ref{assumption:PMatrixEqualEigenValue}}
\label{subsect:RelaxationOfAssumption3}

So far, all the solutions discussed in Section~\ref{subsect:OptimalSolutionForCase1} and \ref{subsect:OptimalSolutionForCase2}
assume Assumption~\ref{assumption:PMatrixEqualEigenValue}. Simply put, the Lyapunov ellipsoid $\mathcal{E}$ should be a hyper sphere. However, in practice, $\mathcal{E}$ usually is not a hyper sphere. Instead, $\mathcal{E}$ usually has unequal principal axes lengths, and the principal axes usually are not parallel to the coordinate axes. 

Fortunately, Assumption~\ref{assumption:PMatrixEqualEigenValue} can be removed by applying linear transformations to the $n$-dimensional state space.

For narrative convenience, let us denote the original $n$-dimensional state space as $\mathbb{S}_{1}$, and its coordinate system as $\mathbb{C}_{1}$. In $\mathbb{S}_{1}$, and \emph{assuming coordinate system} (a.c.s.) $\mathbb{C}_{1}$,  we rewrite everything. 

To start, the original linear control system  \eqref{eqn:LinearSystemModel} becomes 
\begin{eqnarray}
    \begin{cases}
        \dot{\vec{x}}_{1} = A_{1}(\vec{x}_{1} - \vec{x}_{{\sf o},1})+B_{1}\vec{u}_{1}, \\
        \vec{u}_{1} = -K_{1}(\vec{x}_{1} - \vec{x}_{{\sf o},1}), 
    \end{cases} 
    \label{eqn:LinearSystemModelS1C1}
\end{eqnarray}
\noindent where $\vec{x}_{1} \in \mathbb{R}^{n}$ is the plant state, and $\vec{x}_{{\sf o},1} \in \mathcal{R}_{{\sf o},1} \subseteq \mathbb{R}^{n}$ is the given original reference state. Here  $\mathcal{R}_{{\sf o,1}}$ is the feasible region of reference state. Correspondingly, we rewrite Assumption~\ref{assumption:ClosedFeasibleRegionOfTheReferenceState} and \ref{assumption:ConstantReferenceState} respectively as $\mathbb{S}_{1}$-Assumption~\ref{assumption:ClosedFeasibleRegionOfTheReferenceStateS1C1} and \ref{assumption:ConstantReferenceStateS1C1}:
\vspace{0.1in}
\hrule
\begin{S1_assumption}
\label{assumption:ClosedFeasibleRegionOfTheReferenceStateS1C1}
$\mathcal{R}_{{\sf o},1}$ is \emph{closed}, and is defined in $\mathbb{S}_{1}$ (a.c.s. $\mathbb{C}_{1}$) by a set of linear constraints, aka \emph{reference state constraints}, denoted by 
\begin{eqnarray}
    \label{eqn:ReferenceStateConstraintsS1C1}
    g_j(\vec{x}_{{\sf o},1}) \stackrel{{\sf def}}{=} \vec{\omega}_{j,1} \cdot \vec{x}_{{\sf o},1} + b_{j,1} \leqslant 0, \;\; j = 1, 2, \ldots, r. 
\end{eqnarray}
\end{S1_assumption}
\hrule
\vspace{0.1in}
\hrule
\begin{S1_assumption}
\label{assumption:ConstantReferenceStateS1C1}
    Unless otherwise denoted (specifically,  when switching the reference state), we assume $\vec{x}_{{\sf o},1}$ is constant. 
\end{S1_assumption}
\hrule
\vspace{0.1in}
\noindent Also as before, in \eqref{eqn:LinearSystemModelS1C1}, $A_{1} \in \mathbb{R}^{n \times n}$ and $B_{1} \in \mathbb{R}^{n \times m}$ are given as per the physical system, and $K_{1} \in \mathbb{R}^{m \times n}$ is the to-be-designed linear controller.
Correspondingly, the Lyapunov equation \eqref{eqn:LyapunovEquation} becomes
\begin{eqnarray}
A_{{\sf cl},1}^{\sf T}P_{1} + P_{1}A_{{\sf cl},1} = -Q_{1}, \label{eqn:LyapunovEquationS1C1}
\end{eqnarray}
where $A_{{\sf cl},1} \stackrel{{\sf def}}{=} (A_{1} - B_{1}K_{1}) \in \mathbb{R}^{n\times n}$. Suppose through LMI, we get the solution to the above Lyapunov equation: $P_{1}$ and $Q_{1}$ (both as symmetric positive definite $\mathbb{R}^{n \times n}$ matrices), and the linear controller $K_{1}$. 
Correspondingly, we get the Lyapunov ellipsoid $\mathcal{E}_{1}$ as follows: 
\begin{eqnarray}
    \mathcal{E}_{1} &=& E(\vec{x}_{1}(t_0), \vec{x}_{{\sf o},1}, P_{1}), 
    %
    \label{eqn:mathcal_E_1}
\end{eqnarray}
\noindent where $\vec{x}_{1}(t)$ is the plant state at time instance $t$, $t_0$ is the initial time instance, and $E$ is defined in \eqref{eqn:LyapunovEllipsoidFunction}.

Suppose at time $t_{1}$, the original operational region $\bar{\mathcal{F}}_{1}$ changes to the new operational region $\bar{\mathcal{F}}'_{1}$. Correspondingly, we rewrite Assumption~\ref{assumption:LinearConstraints} and \ref{assumption:Rescueable} respectively as $\mathbb{S}_{1}$-Assumption \ref{assumption:LinearConstraintsS1C1} and \ref{assumption:RescueableS1C1}.
\vspace{0.1in}
\hrule
\begin{S1_assumption} 
    \label{assumption:LinearConstraintsS1C1}
    The new operational region $\bar{\mathcal{F}}'_{1}$ is compact (i.e. closed and bounded) and convex, and is defined in $\mathbb{S}_{1}$ (a.c.s. $\mathbb{C}_{1}$) by a set of linear operational constraints:
    \begin{eqnarray}
        \label{eqn:NewReachabilityConstraintsS1C1}
        \vec{v}_{k,1} \cdot \vec{x}_{1} + \beta_{k,1} \leqslant 0,\;\; k=1,2,\ldots, s. 
    \end{eqnarray}
\end{S1_assumption}
\hrule
\vspace{0.1in}
\hrule
\begin{S1_assumption} 
    \label{assumption:RescueableS1C1}
    The present plant state $\vec{x}_{1}(t_1) \in \bar{\mathcal{F}}'_{1}$.
\end{S1_assumption}
\hrule
\vspace{0.1in}

Note in $\mathbb{S}_{1}$, Assumption~\ref{assumption:PMatrixEqualEigenValue} now no longer holds.

\vspace{0.1in}

With the above contexts, at $t_{1}$, to maintain the reachability safety, we aim to find a new reference state $\vec{x}'_{{\sf o},1} \in \mathcal{R}_{{\sf o},1}$ , so that the new linear control system becomes
\begin{eqnarray}
\begin{cases}
\dot{\vec{x}}_{1} = A_{1}(\vec{x}_{1} - \vec{x}'_{{\sf o},1})+B\vec{u}_{1}, \\
\vec{u}_{1} = -K_{1}(\vec{x}_{1} - \vec{x}'_{{\sf o},1}), 
\end{cases} 
\label{eqn:NewLinearSystemModelS1C1}
\end{eqnarray}
\noindent Note \eqref{eqn:NewLinearSystemModelS1C1} is just a rewriting of \eqref{eqn:NewLinearSystemModel}, emphasizing that we are describing the system in $\mathbb{S}_{1}$ (a.c.s. $\mathbb{C}_{1}$).

We demand $\vec{x}'_{{\sf o},1}$ to satisfy the following requirements. 
%
%
\begin{description}
    \item[({\bf $\mathbb{S}_{1}$-R1}):] (Obligatory) Confine the new linear control system \eqref{eqn:NewLinearSystemModelS1C1}'s future trajectory of $\vec{x}_{1}(t)$ ($t \geqslant t_{1}$), denoted as $\{\vec{x}_{1}(t)\}_{t \geqslant t_{1}}$, within a new Lyapunov ellipsoid of the following form
    \begin{eqnarray}
    \label{eqn:NewLyapunovEllipsoidS1C1}
    && \mathcal{E}''_{1} = E(\vec{x}_{1}(t_1), \vec{x}'_{{\sf o},1}, P_{1}) \nonumber \\
    &=& \left\{ \vec{\xi}_{1} \, \big| V_{\vec{x}'_{{\sf o},1}, P_{1}} (\vec{\xi}_{1}) \leqslant V_{\vec{x}'_{{\sf o},1}, P_{1}}(\vec{x}_{1}(t_1)),\;\vec{\xi}_{1} \in \mathbb{R}^n \right\}, 
    \end{eqnarray}
    \noindent where (in compliance with the definition by  \eqref{eqn:LyapunovFunction}) 
    \begin{eqnarray}
    \label{eqn:NewLyapunovPotentialEnergyS1C1}
    V_{\vec{x}'_{{\sf o},1}, P_{1}}(\vec{\xi}_{1}) = (\vec{\xi}_{1} - \vec{x}'_{{\sf o},1})^{{\sf T}} P_{1} (\vec{\xi}_{1} - \vec{x}'_{{\sf o},1}), 
    \end{eqnarray}
    \noindent and $\mathcal{E}''_{1} \cap \mathcal{F}'_{1} = \varnothing$. 
    \item[({\bf $\mathbb{S}_{1}$-R2}):] (Obligatory) Confine $\vec{x}'_{{\sf o},1}$ within the feasible region of the reference state (see  \eqref{eqn:ReferenceStateConstraintsS1C1}), i.e. $\vec{x}'_{{\sf o},1} \in \mathcal{R}_{{\sf o},1}$. 
    \item[({\bf $\mathbb{S}_{1}$-R3}):] (Optional and Heuristic) Minimize the volume of $\mathcal{E}''_{1}$.
\end{description}
\vspace{0.1in}

To find the $\vec{x}'_{{\sf o},1}$ that satisfies ({\bf $\mathbb{S}_{1}$-R1}) $\sim$ ({\bf $\mathbb{S}_{1}$-R3}), we propose to linearly transform the state space $\mathbb{S}_{1}$ (a.c.s. $\mathbb{C}_{1}$) to another state space $\mathbb{S}_{2}$ (a.c.s. $\mathbb{C}_{2}$), to make Assumption~\ref{assumption:PMatrixEqualEigenValue} hold again. Thus, the problem formulation and solution described in Section~\ref{sect:ProblemFormulation}, \ref{subsect:OptimalSolutionForCase1} $\sim$ \ref{subsect:Step3OfThe3StepProcedure} can be reused.

Specifically, we notice that as a solution to the Lyapunov equation~\eqref{eqn:LyapunovEquationS1C1},  $P_{1}$ must be a symmetric positive definite $\mathbb{R}^{n \times n}$ matrix. According to linear algebra~\cite[pp.648]{boyd2004convex}, using the seminal \emph{Singular Value Decomposition} (SVD), $P_{1}$ can always be decomposed to the following form
\begin{eqnarray}
P_{1} = U \Lambda U^{\sf T}, \label{eqn:P_SVD}
\end{eqnarray}
\noindent where $U \in \mathbb{R}^{n \times n}$ is an orthogonal matrix (i.e. $UU^{{\sf T}} = U^{{\sf T}}U = I$, where $I$ is the $\mathbb{R}^{n \times n}$ identity matrix), and $\Lambda \in \mathbb{R}^{n \times n}$ is a diagonal matrix with $P_{1}$'s eigen values as its diagonal elements. Note $P_{1}$ is positive definite, hence every diagonal element of $\Lambda$ is positive. Furthermore, SVD can be conducted in a way so that the diagonal elements of $\Lambda$ are sorted in descending order.

Let ${\sf diag}(e_1, e_2, \ldots, e_n)$ represent a diagonal matrix whose diagonal elements (respectively from row $1$ to $n$) are $e_1$, $e_2$, $\ldots$, $e_n$. We can denote 
\begin{eqnarray}
\Lambda = {{\sf diag}}(\lambda_1, \lambda_2, \ldots, \lambda_n), 
\; \mbox{where} \; \lambda_1 \geqslant \lambda_2 \geqslant \ldots \geqslant \lambda_n > 0; \label{eqn:Lambda}
\end{eqnarray}
\noindent and denote 
\begin{eqnarray}
\Lambda^{-\frac{1}{2}} &\stackrel{{\sf def}}{=}& {{\sf diag}}(\frac{1}{\sqrt{\lambda_1}}, \frac{1}{\sqrt{\lambda_2}}, \ldots, \frac{1}{\sqrt{\lambda_n}}), \label{eqn:Lambda_-frac1_2} \\
\Lambda^{\frac{1}{2}} &\stackrel{{\sf def}}{=}& {{\sf diag}}({\sqrt{\lambda_1}}, {\sqrt{\lambda_2}}, \ldots, {\sqrt{\lambda_n}}). \label{eqn:Lambda_frac1_2}
\end{eqnarray}

Let us carry out the following linear transformation, denoted as $T_{1 \rightarrow 2}$, of all vectors in state space $\mathbb{S}_{1}$ (a.c.s. $\mathbb{C}_{1}$) to state space $\mathbb{S}_{2}$ (and refer to the corresponding coordinate system in $\mathbb{S}_{2}$ as $\mathbb{C}_{2}$).

$\forall \vec{\xi}_{1} \in \mathbb{S}_{1}$, $\vec{\xi}_{1}$ is linearly transformed to $\vec{\xi}_{2} \in \mathbb{S}_{2}$ as per
\begin{eqnarray}
    \vec{\xi}_{2} = T_{1 \rightarrow 2} (\vec{\xi}_{1}) \stackrel{{\sf def}} {=} \Lambda^{\frac{1}{2}}U^{{\sf T}} \vec{\xi}_{1}. \label{eqn:T_1_2}
\end{eqnarray}

Obviously, the inverse transformation, denoted as $T_{2 \rightarrow 1}$, is
$\forall \vec{\xi}_{2} \in \mathbb{S}_{2}$, $\vec{\xi}_{2}$ is linearly transformed to $\vec{\xi}_{1} \in \mathbb{S}_{1}$ as per
\begin{eqnarray}
    \vec{\xi}_{1} = T_{2 \rightarrow 1}(\vec{\xi}_{2}) \stackrel{{\sf def}}{=} U \Lambda^{-\frac{1}{2}} \vec{\xi}_{2}. \label{eqn:T_2_1}
\end{eqnarray}

Meanwhile, we have the following lemma.

\vspace{0.1in}
\hrule
\begin{lemma}[one-to-one mapping of $P_{1}$-based hyper ellipsoid in $\mathbb{S}_{1}$ and hyper sphere in $\mathbb{S}_{2}$]
    \label{lemma:HyperEllipsoidInS1HyperSphereInS2}
    Given a symmetric positive definite matrix $P_{1} \in \mathbb{R}^{n \times n}$ and its SVD as per \eqref{eqn:P_SVD} (which decides the value of $U$ and $\Lambda$, and $\Lambda$ complies with \eqref{eqn:Lambda}). Given any $\vec{\xi}_{{\sf p},1} \in \mathbb{R}^n$ and $\vec{\xi}_{{\sf o}, 1} \in \mathbb{R}^n$. Then a so-called \emph{$P_{1}$-based hyper ellipsoid} in $\mathbb{S}_{1}$ (a.c.s. $\mathbb{C}_{1}$) defined by 
    \begin{eqnarray}
        \mathcal{O}_{1} &\stackrel{{\sf def}}{=}& \left\{ \vec{\xi}_{1} \, \big|  
        (\vec{\xi}_{1} - \vec{\xi}_{{\sf o},1})^{{\sf T}} P_{1} (\vec{\xi}_{1} - \vec{\xi}_{{\sf o},1}) \right. \nonumber \\
        && \left. \leqslant
        (\vec{\xi}_{{\sf p}, 1} - \vec{\xi}_{{\sf o},1})^{{\sf T}} P_{1} (\vec{\xi}_{{\sf p},1} - \vec{\xi}_{{\sf o},1}),
        \;\vec{\xi}_{1} \in \mathbb{R}^n \right\} \nonumber \\
        &=& \left\{ \vec{\xi}_{1} \, \big| V_{\vec{\xi}_{{\sf o},1}, P_{1}} (\vec{\xi}_{1}) \leqslant V_{\vec{\xi}_{{\sf o},1}, P_{1}}(\vec{\xi}_{{\sf p},1}),\;\vec{\xi}_{1} \in \mathbb{R}^n \right\} \;\; \mbox{(see \eqref{eqn:LyapunovFunction})} \nonumber \\
        &=& E(\vec{\xi}_{{\sf p}, 1}, \vec{\xi}_{{\sf o}, 1}, P_{1})  \;\; \mbox{(see \eqref{eqn:LyapunovEllipsoidFunction})}       
        \label{eqn:O_1_details} 
    \end{eqnarray}
    \noindent is translated by linear transformation $T_{1 \rightarrow 2}$ (see \eqref{eqn:T_1_2}) into a hyper sphere in $\mathbb{S}_{2}$ (a.c.s. $\mathbb{C}_{2}$) defined by 
    \begin{eqnarray}
        \mathcal{O}_2 &\stackrel{{\sf def}}{=}& \left\{ \vec{\xi}_{2} \big| 
        ( \vec{\xi}_{2} - \vec{\xi}_{{\sf o},2})^{{\sf T}} (\vec{\xi}_{2} - \vec{\xi}_{{\sf o},2}) \right. \nonumber \\
        && \leqslant
        \left. (\vec{\xi}_{{\sf p},2} - \vec{\xi}_{{\sf o},2})^{{\sf T}}  (\vec{\xi}_{{\sf p},2} - \vec{\xi}_{{\sf o},2}), \vec{\xi}_{2} \in \mathbb{R}^n \right\}, \label{eqn:O_2_details}
    \end{eqnarray}
    \noindent where 
    \begin{eqnarray}
    %
    %
    %
    \vec{\xi}_{{\sf o},2} &=& T_{1 \rightarrow 2}(\vec{\xi}_{{\sf o},1}) = \Lambda^{\frac{1}{2}} U^{{\sf T}} \vec{\xi}_{{\sf o},1} \label{eqn:xi_o_2} \\
    %
    %
    \mbox{and} \;\; \vec{\xi}_{{\sf p},2} &=& T_{1 \rightarrow 2}(\vec{\xi}_{{\sf p},1}) = \Lambda^{\frac{1}{2}} U^{{\sf T}} \vec{\xi}_{{\sf p},1}. \label{eqn:xi_p_2} 
    %
    %
    \end{eqnarray}
    Conversely, given any $\vec{\xi}_{{\sf o},2} \in \mathbb{R}^n$ and $\vec{\xi}_{{\sf p}, 2} \in \mathbb{R}^n$. Then the hyper sphere in $\mathbb{S}_{2}$ (a.c.s. $\mathbb{C}_{2}$) defined by \eqref{eqn:O_2_details} is translated by linear transformation $T_{2 \rightarrow 1}$ (see \eqref{eqn:T_2_1}) into a $P_{1}$-based hyper ellipsoid in $\mathbb{S}_{1}$ (a.c.s. $\mathbb{C}_{1}$) defined by \eqref{eqn:O_1_details}, where $P_{1} \in \mathbb{R}^{n \times n}$ is the symmetric positive definite matrix defined by \eqref{eqn:P_SVD} and 
    \begin{eqnarray}
    %
    %
    %
    %
    \vec{\xi}_{{\sf o},1} &=& T_{2 \rightarrow 1}(\vec{\xi}_{{\sf o},2}) =  U \Lambda^{-\frac{1}{2}} \vec{\xi}_{{\sf o},2} 
    \label{eqn:xi_o_1} \\
    %
    %
    \mbox{and} \;\; \vec{\xi}_{{\sf p},1} &=& T_{2 \rightarrow 1}(\vec{\xi}_{{\sf p},2}) =  U \Lambda^{-\frac{1}{2}} \vec{\xi}_{{\sf p},2}.
    \label{eqn:x_p_1}
    \end{eqnarray}
\end{lemma}
\hrule
\vspace{0.1in}

\begin{proof}

First, let's prove any $P_{1}$-based hyper ellipsoid $\mathcal{O}_{1}$ in $\mathbb{S}_{1}$ (a.c.s. $\mathbb{C}_{1}$) maps to a hyper sphere $\mathcal{O}_{2}$ in $\mathbb{S}_{2}$ (a.c.s. $\mathbb{C}_{2}$).

Combining \eqref{eqn:T_1_2}\eqref{eqn:xi_o_2}\eqref{eqn:xi_p_2} and \eqref{eqn:O_1_details}, we derive the $T_{1 \rightarrow 2}$ transformed $\mathcal{O}_{1}$ in $\mathbb{S}_{2}$ (a.c.s. $\mathbb{C}_{2}$), denoted as $\mathcal{O}_{2}$, as follows:
\begin{eqnarray}
&& \mathcal{O}_{2} \nonumber \\
&=& \left\{ \vec{\xi}_{2} \big| 
(U \Lambda^{-\frac{1}{2}} \vec{\xi}_{2} - U \Lambda^{-\frac{1}{2}}\vec{\xi}_{{\sf o},2})^{{\sf T}} P_{1} (U \Lambda^{-\frac{1}{2}}\vec{\xi}_{2} - U \Lambda^{-\frac{1}{2}}\vec{\xi}_{{\sf o},2}) \right. \nonumber \\
&& \leqslant
(U \Lambda^{-\frac{1}{2}} \vec{\xi}_{{\sf p},2} - U \Lambda^{-\frac{1}{2}} \vec{\xi}_{{\sf o},2})^{{\sf T}} P_{1} (U \Lambda^{-\frac{1}{2}} \vec{\xi}_{{\sf p},2} - U \Lambda^{-\frac{1}{2}}\vec{\xi}_{{\sf o},2}), \nonumber \\
&& \left. \vec{\xi}_{2} \in \mathbb{R}^n \right\} \nonumber \\
&=& \left\{ \vec{\xi}_{2} \big| 
( \vec{\xi}_{2} - \vec{\xi}_{{\sf o},2})^{{\sf T}} \Lambda^{-\frac{1}{2}} U^{{\sf T}} P_{1} U \Lambda^{-\frac{1}{2}} (\vec{\xi}_{2} - \vec{\xi}_{{\sf o},2}) \right. \nonumber \\
&& \leqslant
\left. (\vec{\xi}_{{\sf p},2} - \vec{\xi}_{{\sf o},2})^{{\sf T}} \Lambda^{-\frac{1}{2}} U^{{\sf T}} P_{1} U \Lambda^{-\frac{1}{2}} (\vec{\xi}_{{\sf p},2} - \vec{\xi}_{{\sf o},2}), \vec{\xi}_{2} \in \mathbb{R}^n \right\} \nonumber \\
&=& \left\{ \vec{\xi}_{2} \big| 
( \vec{\xi}_{2} - \vec{\xi}_{{\sf o},2})^{{\sf T}} (\vec{\xi}_{2} - \vec{\xi}_{{\sf o},2}) \right. \nonumber \\
&& \leqslant
\left. (\vec{\xi}_{{\sf p},2} - \vec{\xi}_{{\sf o},2})^{{\sf T}}  (\vec{\xi}_{{\sf p},2} - \vec{\xi}_{{\sf o},2}), \vec{\xi}_{2} \in \mathbb{R}^n \right\}. \;\; \mbox{(due to \eqref{eqn:P_SVD})} \nonumber \\
\label{eqn:O_2}
\end{eqnarray}

From \eqref{eqn:O_2}, we see $\mathcal{O}_{2}$ is a hyper sphere centered at $\vec{x}_{{\sf o},2}$, with $\vec{x}_{{\sf p}, 2}$ residing on its surface (i.e. with a radius length of $\|\vec{x}_{{\sf p},2} - \vec{x}_{{\sf o},2}\|_2$).  

Next, let's prove any hyper sphere $\mathcal{E}_{2}$ in $\mathbb{S}_{2}$ (a.c.s. $\mathbb{C}_{2}$) maps to a $P_{1}$-based hyper ellipsoid $\mathcal{E}_{1}$ in $\mathbb{S}_{1}$ (a.c.s. $\mathbb{C}_{1}$).

Combining \eqref{eqn:T_2_1}\eqref{eqn:xi_o_1}\eqref{eqn:x_p_1} and \eqref{eqn:O_2_details}, we derive the $T_{2 \rightarrow 1}$ transformed $\mathcal{E}_{2}$ in $\mathbb{S}_{1}$ (a.c.s. $\mathbb{C}_{1}$), denoted as $\mathcal{E}_{1}$, as follows:
\begin{eqnarray}
&& \mathcal{E}_{1} \nonumber \\
&=& \left\{ \vec{\xi}_{1} \big| 
( \Lambda^{\frac{1}{2}}U^{{\sf T}} \vec{\xi}_{1} - \Lambda^{\frac{1}{2}}U^{{\sf T}} \vec{x}_{{\sf o},1})^{{\sf T}} (\Lambda^{\frac{1}{2}}U^{{\sf T}} \vec{\xi}_{1} - \Lambda^{\frac{1}{2}}U^{{\sf T}} \vec{x}_{{\sf o},1}) \right. \nonumber \\
&& \leqslant
(\Lambda^{\frac{1}{2}}U^{{\sf T}} \vec{x}_{{\sf p},1} - \Lambda^{\frac{1}{2}}U^{{\sf T}} \vec{x}_{{\sf o},1})^{{\sf T}}  (\Lambda^{\frac{1}{2}}U^{{\sf T}} \vec{x}_{{\sf p},1} - \Lambda^{\frac{1}{2}}U^{{\sf T}} \vec{x}_{{\sf o},1}), \nonumber \\
&& \left. \vec{\xi}_{1} \in \mathbb{R}^n \right\} \nonumber \\
&=& \left\{ \vec{\xi}_{1} \big| 
( \vec{\xi}_{1} - \vec{x}_{{\sf o},1})^{{\sf T}} U \Lambda^{\frac{1}{2}} \Lambda^{\frac{1}{2}} U^{{\sf T}} (\vec{\xi}_{1} - \vec{x}_{{\sf o},1}) \right. \nonumber \\
&& \leqslant
\left. (\vec{x}_{{\sf p},1} - \vec{x}_{{\sf o},1})^{{\sf T}} U \Lambda^{\frac{1}{2}} \Lambda^{\frac{1}{2}} U^{{\sf T}} (\vec{x}_{{\sf p},1} - \vec{x}_{{\sf o},1}), \vec{\xi}_{1} \in \mathbb{R}^n \right\} \nonumber \\
&=& \left\{ \vec{\xi}_{1} \big| 
( \vec{\xi}_{1} - \vec{x}_{{\sf o},1})^{{\sf T}} P_{1} (\vec{\xi}_{1} - \vec{x}_{{\sf o},1}) \right. \nonumber \\
&& \leqslant
\left. (\vec{x}_{{\sf p},1} - \vec{x}_{{\sf o},1})^{{\sf T}} P_{1} (\vec{x}_{{\sf p},1} - \vec{x}_{{\sf o},1}), \vec{\xi}_{1} \in \mathbb{R}^n \right\}. \;\; \mbox{(due to \eqref{eqn:P_SVD})} \nonumber \\
\label{eqn:E_1}
\end{eqnarray}

From \eqref{eqn:E_1}, we see $\mathcal{E}_{1}$ is a $P_{1}$-based hyper ellipsoid centered at $\vec{x}_{{\sf o},1}$,  with $\vec{x}_{{\sf p}, 1}$ residing on its surface. 
%
%
\end{proof}
\vspace{0.1in}

With the above knowledge in mind, we linearly transform everything of $\mathbb{S}_{1}$ (a.c.s. $\mathbb{C}_{1}$) to $\mathbb{S}_{2}$ (a.c.s. $\mathbb{C}_{2}$). 

Specifically, the new linear control system defined by~\eqref{eqn:NewLinearSystemModelS1C1} in $\mathbb{S}_{1}$ (a.c.s. $\mathbb{C}_{1}$) becomes defined by~\eqref{eqn:NewLinearSystemModelS2C2} in $\mathbb{S}_{2}$ (a.c.s. $\mathbb{C}_{2}$).
\begin{eqnarray}
\begin{cases}
\dot{\vec{x}}_{2} = A_{2}(\vec{x}_{2} - \vec{x}'_{{\sf o},2})+B_{2}\vec{u}_{2}, \\
\vec{u}_{2} = -K_{2}(\vec{x}_{2} - \vec{x}'_{{\sf o},2}), 
\end{cases} 
\label{eqn:NewLinearSystemModelS2C2}
\end{eqnarray}
\noindent where
\begin{eqnarray}
\vec{x}_{2} &=& T_{1\rightarrow2}(\vec{x}_{1}) = \Lambda^{\frac{1}{2}} U^{{\sf T}} \vec{x}_{1}, \nonumber \\
\vec{x}'_{{\sf o},2} &=& T_{1\rightarrow2}(\vec{x}'_{{\sf o},1}) = \Lambda^{\frac{1}{2}} U^{{\sf T}} \vec{x}'_{{\sf o},1}, \nonumber \\
A_{2} &=& \Lambda^{\frac{1}{2}} U^{{\sf T}} A_{1} U \Lambda^{-\frac{1}{2}}, \nonumber \\
B_{2} &=& \Lambda^{\frac{1}{2}} U^{{\sf T}} B_{1} U \Lambda^{-\frac{1}{2}}, \nonumber \\
K_{2} &=& \Lambda^{\frac{1}{2}} U^{{\sf T}} K_{1} U \Lambda^{-\frac{1}{2}}, \label{eqn:NewLinearSystemModelSymbolsS2C2}
\end{eqnarray}

Correspondingly, 
$\mathbb{S}_{1}$-Assumption~\ref{assumption:ClosedFeasibleRegionOfTheReferenceStateS1C1} on the feasible region of reference state $\mathcal{R}_{{\sf o},1}$ in $\mathbb{S}_{1}$ (a.c.s. $\mathbb{C}_{1}$) becomes $\mathbb{S}_{2}$-Assumption~\ref{assumption:ClosedFeasibleRegionOfTheReferenceStateS2C2} on the feasible region of reference state $\mathcal{R}_{{\sf o},2}$ in $\mathbb{S}_{2}$ (a.c.s. $\mathbb{C}_{2}$).
\vspace{0.1in}
\hrule
\begin{S2_assumption}
\label{assumption:ClosedFeasibleRegionOfTheReferenceStateS2C2}
$\mathcal{R}_{{\sf o},2}$ is \emph{closed}, and is defined in $\mathbb{S}_{2}$ (a.c.s. $\mathbb{C}_{2}$) by a set of linear constraints, aka \emph{reference state constraints}, denoted by 
\begin{eqnarray}
    \label{eqn:ReferenceStateConstraintsS2C2}
    g_{j,2}(\vec{x}_{{\sf o},2}) &\stackrel{{\sf def}}{=}& \vec{\omega}_{j,2} \cdot \vec{x}_{{\sf o},2} + b_{j,2} \leqslant 0,  \nonumber \\
    \mbox{where } \vec{\omega}_{j,2} &=& \Lambda^{-\frac{1}{2}} U^{{\sf T}} \vec{\omega}_{j,1}, \nonumber \\
    b_{j,2} &=& b_{j,1}, \qquad\qquad j = 1, 2, \ldots, r.
\end{eqnarray}
\end{S2_assumption}
\hrule
\vspace{0.1in}

$\mathbb{S}_{1}$-Assumption~\ref{assumption:ConstantReferenceStateS1C1} becomes $\mathbb{S}_{2}$-Assumption~\ref{assumption:ConstantReferenceStateS2C2}.
\vspace{0.1in}
\hrule
\begin{S2_assumption}
\label{assumption:ConstantReferenceStateS2C2}
    Unless otherwise denoted (specifically,  when switching the reference state), we assume $\vec{x}_{{\sf o},2}$ is constant. 
\end{S2_assumption}
\hrule
\vspace{0.1in}

Corresponding to the change of the original operational region $\bar{\mathcal{F}}_{1}$ to the new operational region $\bar{\mathcal{F}}'_{1}$ in $\mathbb{S}_{1}$ (a.c.s. $\mathbb{C}_{1}$) at $t_{1}$, we have the change of the original operational region $\bar{\mathcal{F}}_{2}$ to the new operational region $\bar{\mathcal{F}}'_{2}$ in $\mathbb{S}_{2}$ (a.c.s. $\mathbb{C}_{2}$) at $t_{1}$. Correspondingly, $\mathbb{S}_{1}$-Assumption~\ref{assumption:LinearConstraintsS1C1} and \ref{assumption:RescueableS1C1} become $\mathbb{S}_{2}$-Assumption~\ref{assumption:LinearConstraintsS2C2} and \ref{assumption:RescueableS2C2}.
\vspace{0.1in}
\hrule
\begin{S2_assumption} 
\label{assumption:LinearConstraintsS2C2}
The new operational region $\bar{\mathcal{F}}'_{2}$ is compact (i.e. closed and bounded) and convex, and is defined in $\mathbb{S}_{2}$ (a.c.s. $\mathbb{C}_{2}$) by a set of linear operational constraints:
\begin{eqnarray}
\label{eqn:NewReachabilityConstraintsS2C2}
\vec{v}_{k,2} \cdot \vec{x}_{2} + \beta_{k,2} &\leqslant& 0, \nonumber \\
\mbox{where } \vec{v}_{k,2} &=& \Lambda^{-\frac{1}{2}} U^{{\sf T}} \vec{v}_{k,1}, \nonumber \\
\beta_{k,2} &=& \beta_{k,1}, \qquad\quad k=1,2,\ldots, s.
\end{eqnarray}
\end{S2_assumption}
\hrule
\vspace{0.1in}
\hrule
\begin{S2_assumption} 
    \label{assumption:RescueableS2C2}
    The present plant state $\vec{x}_{2}(t_1) \in \bar{\mathcal{F}}'_{2}$.
\end{S2_assumption}
\hrule
\vspace{0.1in}

Next, we shall prove Assumption~\ref{assumption:PMatrixEqualEigenValue} is recovered in $\mathbb{S}_{2}$.

\vspace{0.1in}

In $\mathbb{S}_{2}$ (a.c.s. $\mathbb{C}_{2}$), the  Lyapunov function~\eqref{eqn:LyapunovFunction} of linear control system~\eqref{eqn:NewLinearSystemModelS2C2} becomes
\begin{eqnarray}
& A_{{\sf cl},2}^{\sf T}P_{2} + P_{2}A_{{\sf cl},2} = -Q_{2}, \label{eqn:LyapunovEquationS2C2} \\
& \mbox{where } A_{{\sf cl},2} \stackrel{{\sf def}}{=} (A_{2} - B_{2}K_{2}). \label{eqn:A_cl_2}
\end{eqnarray}

Next we prove 
\vspace{0.1in}
\hrule
\begin{lemma}
\label{lemma:SolutionToLyapunovEquationS2C2}
The following is a solution to the Lyapunov equation~\eqref{eqn:LyapunovEquationS2C2}:
\begin{eqnarray}
P_{2} = I \in \mathbb{R}^{n \times n}, \quad Q_{2} = \Lambda^{-\frac{1}{2}} U^{{\sf T}} Q_{1} U \Lambda^{-\frac{1}{2}}. \label{eqn:P_2_Q_2}
\end{eqnarray}
\end{lemma}
\hrule
\vspace{0.1in}

\begin{proof}

First, obviously $P_{2} = I$ is a symmetric real positive definite matrix. Meanwhile, as
\begin{eqnarray}
&& Q^{{\sf T}} = \Lambda^{-\frac{1}{2}} U^{{\sf T}} Q_{1}^{{\sf T}} U \Lambda^{-\frac{1}{2}} \nonumber \\
&=& \Lambda^{-\frac{1}{2}} U^{{\sf T}} Q_{1} U \Lambda^{-\frac{1}{2}} = Q_{2}, \quad \mbox{($Q_{1}$ is symmetric)}
\end{eqnarray}
\noindent and as $\forall \vec{x}_{2} \in \mathbb{R}^{n}$
\begin{eqnarray}
&& \vec{x}_{2}^{{\sf T}} Q_{2} \vec{x}_{2} = \vec{x}_{2}^{{\sf T}} \Lambda^{-\frac{1}{2}} U^{{\sf T}} Q_{1} U \Lambda^{-\frac{1}{2}} \vec{x}_{2} \nonumber \\
&=& \vec{x}_{1}^{{\sf T}} Q_{1} \vec{x}_{1} \quad \mbox{(see \eqref{eqn:T_2_1})} \nonumber \\
&>& 0, \qquad \mbox{($Q_{1}$ is positive definite)} \nonumber
\end{eqnarray}
\noindent we know $Q_{2} \in \mathbb{R}^{n \times n}$ is also a symmetric real positive definite matrix.

Next, let us prove $P_{2}$, $Q_{2}$ satisfy \eqref{eqn:LyapunovEquationS2C2}. 

Due to \eqref{eqn:LyapunovEquationS1C1} and \eqref{eqn:P_2_Q_2}, we have
\begin{eqnarray}
&& A_{{\sf cl},1}^{{\sf T}} P_{1} + P_{1} A_{{\sf cl},1} = - U \Lambda^{\frac{1}{2}} Q_{2} \Lambda^{\frac{1}{2}} U^{{\sf T}} \nonumber \\
&\Leftrightarrow& \Lambda^{-\frac{1}{2}} U^{{\sf T}} (A_{{\sf cl},1}^{{\sf T}} P_{1} + P_{1} A_{{\sf cl},1}) U \Lambda^{-\frac{1}{2}} = -Q_{2} \nonumber \\
&\Leftrightarrow& \Lambda^{-\frac{1}{2}} U^{{\sf T}} A_{{\sf cl},1}^{{\sf T}} P_{1} U \Lambda^{-\frac{1}{2}} + \Lambda^{-\frac{1}{2}} U^{{\sf T}} P_{1} A_{{\sf cl},1} U \Lambda^{-\frac{1}{2}} = -Q_{2} \nonumber \\
&\Leftrightarrow& \Lambda^{-\frac{1}{2}} U^{{\sf T}} A_{{\sf cl},1}^{{\sf T}} U \Lambda U^{{\sf T}} U \Lambda^{-\frac{1}{2}} + \Lambda^{-\frac{1}{2}} U^{{\sf T}} U \Lambda U^{{\sf T}} A_{{\sf cl},1} U \Lambda^{-\frac{1}{2}} \nonumber \\
&& = -Q_{2}, \qquad\qquad\qquad\qquad\qquad\qquad \mbox{(due to \eqref{eqn:P_SVD})} \nonumber \\
&\Leftrightarrow& \Lambda^{-\frac{1}{2}} U^{{\sf T}} A_{{\sf cl},1}^{{\sf T}} U \Lambda^{\frac{1}{2}} + \Lambda^{\frac{1}{2}} U^{{\sf T}} A_{{\sf cl},1} U \Lambda^{-\frac{1}{2}} = -Q_{2}. \label{eqn:LyapunovEquationA_cl_1_Q_2}
\end{eqnarray}
\noindent Meanwhile \eqref{eqn:A_cl_2}\eqref{eqn:NewLinearSystemModelSymbolsS2C2} implies
\begin{eqnarray}
A_{{\sf cl},2} &=& \Lambda^{\frac{1}{2}} U^{{\sf T}} A_{1} U \Lambda^{-\frac{1}{2}} - \Lambda^{\frac{1}{2}} U^{{\sf T}} B_{1} U \Lambda^{-\frac{1}{2}} \Lambda^{\frac{1}{2}} U^{{\sf T}} K_{1} U \Lambda^{-\frac{1}{2}} \nonumber \\
&=& \Lambda^{\frac{1}{2}} U^{{\sf T}} A_{1} U \Lambda^{-\frac{1}{2}} - \Lambda^{\frac{1}{2}} U^{{\sf T}} B_{1} K_{1} U \Lambda^{-\frac{1}{2}} \nonumber \\
&=& \Lambda^{\frac{1}{2}} U^{{\sf T}} (A_{1} - B_{1} K_{1}) U \Lambda^{-\frac{1}{2}} \nonumber \\
&=& \Lambda^{\frac{1}{2}} U^{{\sf T}} A_{{\sf cl},1} U \Lambda^{-\frac{1}{2}}. \label{eqn:A_cl_2_A_cl_1_Relationship}
\end{eqnarray}
\noindent Therefore, \eqref{eqn:LyapunovEquationA_cl_1_Q_2}\eqref{eqn:A_cl_2_A_cl_1_Relationship} implies $A_{{\sf cl},2}^{{\sf T}} + A_{{\sf cl},2} = -Q_{2}$, i.e. 
\begin{eqnarray}
A_{{\sf cl},2}^{{\sf T}} I + I A_{{\sf cl},2} = -Q_{2}. \label{eqn:I_and_Q_2_AreTheSolution}
\end{eqnarray}
\noindent Therefore, $P_{2} = I$ and $Q_{2}$ are the solution to the Lyapunov equation~\eqref{eqn:LyapunovEquationS2C2}. 

%
\end{proof}
\vspace{0.1in}

Lemma~\ref{lemma:SolutionToLyapunovEquationS2C2} implies that Assumption~\ref{assumption:PMatrixEqualEigenValue}
still holds for the new linear control system~\eqref{eqn:NewLinearSystemModelS2C2} in $\mathbb{S}_{2}$ (a.c.s. $\mathbb{C}_{2}$). Let us rewrite it as
\vspace{0.1in}
\hrule
\begin{S2_assumption} 
    \label{assumption:PMatrixEqualEigenValueS2C2}
    All the eigenvalues of $P_{2}$ have a same positive real value. 
\end{S2_assumption}
\hrule
\vspace{0.1in}

The requirements {\bf $\mathbb{S}_{1}$-R1} $\sim$ {\bf $\mathbb{S}_{1}$-R3} are also rewritten. Specifically, at $t_{1}$, to maintain the reachability safety, we aim to find a new reference state $\vec{x}'_{{\sf o},2} \in \mathcal{R}_{{\sf o},2}$ to satisfy the following requirements. 
%
%
\begin{description}
    \item[({\bf $\mathbb{S}_{2}$-R1}):] (Obligatory) Confine the new linear control system \eqref{eqn:NewLinearSystemModelS2C2}'s future trajectory of $\vec{x}_{2}(t)$ ($t \geqslant t_{1}$), denoted as $\{\vec{x}_{2}(t)\}_{t \geqslant t_{1}}$, within a new Lyapunov ellipsoid of the following form
    \begin{eqnarray}
    \label{eqn:NewLyapunovEllipsoidS2C2}
    && \mathcal{E}''_{2} = E(\vec{x}_{2}(t_1), \vec{x}'_{{\sf o},2}, P_{2}) \nonumber \\
    &=& \left\{ \vec{\xi}_{2} \, \big| V_{\vec{x}'_{{\sf o},2}, P_{2}} (\vec{\xi}_{2}) \leqslant V_{\vec{x}'_{{\sf o},2}, P_{2}}(\vec{x}_{2}(t_1)),\;\vec{\xi}_{2} \in \mathbb{R}^n \right\}, 
    \end{eqnarray}
    \noindent where (in compliance with the definition by  \eqref{eqn:LyapunovFunction}) 
    \begin{eqnarray}
    \label{eqn:NewLyapunovPotentialEnergyS2C2}
    V_{\vec{x}'_{{\sf o},2}, P_{2}}(\vec{\xi}_{2}) = (\vec{\xi}_{2} - \vec{x}'_{{\sf o},2})^{{\sf T}} P_{2} (\vec{\xi}_{2} - \vec{x}'_{{\sf o},2}), 
    \end{eqnarray}
    \noindent and $\mathcal{E}''_{2} \cap \mathcal{F}'_{2} = \varnothing$. 
    \item[({\bf $\mathbb{S}_{2}$-R2}):] (Obligatory) Confine $\vec{x}'_{{\sf o},2}$ within the feasible region of the reference state (see  \eqref{eqn:ReferenceStateConstraintsS2C2}), i.e. $\vec{x}'_{{\sf o},1} \in \mathcal{R}_{{\sf o},2}$. 
    \item[({\bf $\mathbb{S}_{2}$-R3}):] (Optional and Heuristic) Minimize the volume of $\mathcal{E}''_{2}$.
\end{description}
\vspace{0.1in}

Now, because $\mathbb{S}_{2}$-Assumption~\ref{assumption:ClosedFeasibleRegionOfTheReferenceStateS2C2} $\sim$ \ref{assumption:PMatrixEqualEigenValueS2C2} all hold, we can reuse the method described in Section~\ref{sect:ProblemFormulation}, \ref{subsect:OptimalSolutionForCase1} $\sim$ \ref{subsect:Step3OfThe3StepProcedure}, i.e. use the ORSOP problem (see Problem~\ref{Problem_I}), to model and solve our problem: find the $\vec{x}'^{*}_{{\sf o},2}$ that satisfies {\bf $\mathbb{S}_{2}$-R1} $\sim$ {\bf $\mathbb{S}_{2}$-R3} in $\mathbb{S}_{2}$ for the new linear control system~\eqref{eqn:NewLinearSystemModelS2C2}. 

Once the $\vec{x}'^{*}_{{\sf o},2}$ is found, then we can get its mapping in $\mathbb{S}_{1}$ (a.c.s. $\mathbb{C}_{1}$) with the inverse linear transformation~\eqref{eqn:T_2_1}:
\begin{eqnarray}
\vec{x}'^{*}_{{\sf o},1} = T_{2 \rightarrow 1}(\vec{x}'^{*}_{{\sf o},2}) = U \Lambda^{-\frac{1}{2}} \vec{x}'^{*}_{{\sf o},2}. \label{eqn:x'*_o1}
\end{eqnarray}

\balance

We have the following theorem: 
\vspace{0.1in}
%
%
\hrule width 0.489\textwidth \relax
\begin{theorem}
\label{theorem:x_prime_asterisk_o1solvesS1R1-S1R3}
The $\vec{x}'^{*}_{{\sf o},1}$ derived from \eqref{eqn:x'*_o1} is the solution for {\bf $\mathbb{S}_{1}$-R1} $\sim$ {\bf $\mathbb{S}_{1}$-R3} for the new linear control system~\eqref{eqn:NewLinearSystemModelS1C1} in $\mathbb{S}_{1}$ (a.c.s. $\mathbb{C}_{1}$).
\end{theorem}
%
%
\hrule width 0.489\textwidth \relax
\vspace{0.1in}

\begin{proof}

Beause $P_{2} = I$ (see Lemma~\ref{lemma:SolutionToLyapunovEquationS2C2}),  \eqref{eqn:NewLyapunovEllipsoidS2C2} can be rewritten as
\begin{eqnarray}
&& \mathcal{E}''_{2} = E(\vec{x}_{2}(t_1), \vec{x}'_{{\sf o},2}, P_{2}) \nonumber \\
&=& \left\{ \vec{\xi}_{2} \, \big| V_{\vec{x}'_{{\sf o},2}, P_{2}} (\vec{\xi}_{2}) \leqslant V_{\vec{x}'_{{\sf o},2}, P_{2}}(\vec{x}_{2}(t_1)),\;\vec{\xi}_{2} \in \mathbb{R}^n \right\} \nonumber \\
&=& \left\{ \vec{\xi}_{2} \, \big| (\vec{\xi}_{2} - \vec{x}'_{{\sf o},2})^{{\sf T}} (\vec{\xi}_{2} - \vec{x}'_{{\sf o},2}) \right. \nonumber \\
&& \left. \leqslant (\vec{x}_{2}(t_{1}) - \vec{x}'_{{\sf o},2})^{{\sf T}} (\vec{x}_{2}(t_{1}) - \vec{x}'_{{\sf o},2}), \,  \vec{\xi}_{2} \in \mathbb{R}^n \right\}. \label{eqn:NewLyapunovEllipsoidS2C2Details}
\end{eqnarray}

Due to \eqref{eqn:NewLyapunovEllipsoidS2C2Details} and Lemma~\ref{lemma:HyperEllipsoidInS1HyperSphereInS2},  $\vec{x}'^{*}_{{\sf o},2}$'s compliance with {\bf $\mathbb{S}_{2}$-R1} implies $\vec{x}'^{*}_{{\sf o},1}$'s compliance with {\bf $\mathbb{S}_{1}$-R1}. $\hfill(\star)$

As $\mathcal{R}_{{\sf o},1}$ and $\vec{x}'^{*}_{{\sf o},1}$ in $\mathbb{S}_{1}$ (a.c.s. $\mathbb{C}_{1}$) is linearly mapped with $\mathcal{R}_{{\sf o},2}$ and $\vec{x}'^{*}_{{\sf o},2}$ in $\mathbb{S}_{2}$ (a.c.s. $\mathbb{C}_{2}$), $\vec{x}'^{*}_{{\sf o},2}$'s compliance with {\bf $\mathbb{S}_{2}$-R2} implies $\vec{x}'^{*}_{{\sf o},1}$'s compliance with {\bf $\mathbb{S}_{1}$-R2}. $\hfill(\dagger)$

Due to \eqref{eqn:NewLyapunovEllipsoidS2C2Details} and Lemma~\ref{lemma:HyperEllipsoidInS1HyperSphereInS2},  $\vec{x}'^{*}_{{\sf o},2}$'s compliance with {\bf $\mathbb{S}_{2}$-R3} implies $\vec{x}'^{*}_{{\sf o},1}$'s compliance with {\bf $\mathbb{S}_{1}$-R3}. For otherwise, the existence of a better solution in $\mathbb{S}_{1}$ will map to a better solution in $\mathbb{S}_{2}$. $\hfill(\ddagger)$

Combining $(\star)(\dagger)(\ddagger)$, the theorem is proved. 

%
%
\end{proof}
\vspace{0.1in}



\bibliographystyle{IEEEtran}
\bibliography{main}



\end{document}